\documentclass[10pt]{article}
\usepackage{geometry}
\usepackage{titling}
\usepackage{authblk}
\usepackage{amssymb}
\usepackage{amsmath}
\usepackage{amsthm}
\allowdisplaybreaks[4]
\usepackage{slashed}
\usepackage{graphicx,color,epsfig}
\usepackage{multirow}
\usepackage{cite}
\usepackage[colorlinks=true,linkcolor=red,citecolor=blue]{hyperref}
\def\cbar{\overline{c}}

\begin{document}
\title{Investigation of  Effects of New Physics in $\Lambda_b\to\Lambda_c \tau\bar\nu_\tau$  Decay}
\author[1]{Xiao-Long Mu}
\author[1,2]{Ying Li$\footnote{liying@ytu.edu.cn}$}
\author[1]{Zhi-Tian Zou}
\author[1]{Bin Zhu}
\affil[1]{\it Department of Physics, Yantai University, Yantai 264005, China}
\affil[2]{\it Center for High Energy Physics, Peking University, Beijing 100871, China}
\maketitle
\vspace{0.2cm}

\begin{abstract}
Recent experimental results of ${\cal R}(D^{(*)})$ deviate from the standard model (SM) by $3.1\sigma$, suggesting a new physics (NP) that affects the $b\to c \tau \bar\nu_\tau$ transition. Motivated by this, we investigate the possible NP effects in the $\Lambda_b\to\Lambda_c \tau\bar\nu_\tau$ decay. For this purpose, assuming the neutrinos are left-handed, we calculate in detail the helicity amplitudes of $\Lambda_b\to\Lambda_c \ell\bar\nu_\ell$ ($\ell=e^-,\mu^-$ and $\tau^-$) decays with all possible four-fermion operators. Within the latest results of $\Lambda_b\to\Lambda_c$ form factors from lattice QCD calculations, we study these decays in a model-independent manner. The differential and total branching fractions, the longitudinal polarizations of final leptons and hadrons, the forward-backward asymmetries in the lepton-side, the convexity parameters, and the ratio ${\cal R}(\Lambda_c)$ are calculated. In SM, we obtain the ratio ${\cal R}(\Lambda_c)=0.33\pm0.01$. Supposing that NP only affects the third generation fermions, we present the correlations among ${\cal R}(D)$,  ${\cal R}(D^*)$ and ${\cal R}(\Lambda_c)$, as the $\Lambda_b\to\Lambda_c \tau\bar\nu_\tau$ and $B\to D^{(*)}\tau\bar\nu_\tau$ are all induced by $b\to c \ell\bar\nu_\ell$. We perform a minimum $\chi^2$ fit of the wilson coefficient of each operator to the latest experimental data of different observables, including the ratios ${\cal R}(D^{(*)})$ and ${\cal R}(J/\psi)$ and the $\tau$ polarization $P_\tau(D^*)$. It is found that the left-handed scalar operator ${\cal O}_{SL}$ affects the branching fraction remarkably, and the ratio ${\cal R}(\Lambda_c)$ can be enhanced by $30\%$. For other operators, the ratio amounts to $0.38\pm0.02$, which is larger than prediction of SM by $20\%$.  Using the fitted values of the wilson coefficients of the single NP operators, we also give a prognosis for the physical observables of $\Lambda_b\to\Lambda_c \tau\bar\nu_\tau$, including the ratio ${\cal R}(\Lambda_c)$, forward-backward asymmetry and other polarized observables as well as the differential branching fraction. Furthermore, we also study the effects of three typical NP models on the ratio and the differential branching fraction of $\Lambda_b\to\Lambda_c \tau\bar\nu_\tau$.  We hope our results can be tested in the current LHCb experiment and the future high energy experiments.

\end{abstract}

\newpage
\section{Introduction}
After the discovery of Higgs boson, one of the most important tasks is searching for the possible new physics (NP) beyond the standard model (SM). So far, at the LHC people has not directly observed any signals of new particles beyond SM. However, there are currently a number of anomalies in semi-leptonic $b$-flavor hadrons decays, which have attracted a great attention in the scientific community. The unexpected deviations seem to appear in both $b\to c$ and $b\to s$ semi-leptonic decay transitions when different generations of leptons are involved, see refs. \cite{Bifani:2018zmi,Li:2018lxi,Kumbhakar:2019avh} for recent reviews. Unlike the flavor-changing neutral-current $b\to s$ transition that has been thought of as an ideal plate for probing the effect of NP, $b\to c$ transition is a tree level process by the exchange of a $W$ boson in SM. Primordially, in order to study the charged Higgs contribution, the ratios are defined as \cite{Korner:1989qb,Tanaka:1994ay}:
\begin{eqnarray}
{\cal R}(D^{(*)})=\frac{\mathcal{B}(B\to D^{(*)} \tau\bar \nu)}{\mathcal{B}(B\to D^{(*)} \ell\bar \nu)}, \ell=e,\mu.
\end{eqnarray}
For these two ratios, the uncertainties from the Cabibbo-Kobayashi-Maskawa (CKM) matrix element $V_{cb}$ can be removed, and ones from the hadronic transition form factors can also be reduced largely. In SM, based on the heavy-quark effective theory or the lattice QCD approach, they have been studied explicitly \cite{Aoki:2016frl, Fajfer:2012vx, Bigi:2016mdz, Bernlochner:2017jka, Bigi:2017jbd, Jaiswal:2017rve}, and the theoretical averaged results are given by \cite {Amhis:2016xyh}
\begin{eqnarray}
{\cal R}(D)^{\mathrm{SM}}=0.299^{+0.003}_{-0.003}, \quad  \text{  and  } \quad
{\cal R}(D^{*})^{\mathrm{SM}}=0.258^{+0.005}_{-0.005}.
\end{eqnarray}
However, from 2012, measurements from BaBar \cite{Lees:2012xj, Lees:2013uzd}, Belle \cite{Huschle:2015rga, Sato:2016svk, Hirose:2016wfn} and LHCb \cite{Aaij:2015yra, Aaij:2017uff} showed that there are deviations between experimental data and predictions of SM, and the combination results performed by the Heavy Flavour Averaging Group (HFLAV) \cite{Amhis:2016xyh} are
\begin{eqnarray}
{\cal R}(D)^{\text{avg}}    = 0.407 \pm 0.039 \pm 0.024  \quad   \text{  and  }  \quad
{\cal R}(D^{*})^{\text{avg}}= 0.306 \pm 0.013 \pm 0.007\, ,
\end{eqnarray}
with a correlation of $-20\%$, which shows a tension of $3.8\sigma$ with SM predictions. Very recently, at Moriond EW 2019, the Belle collaboration released  the latest measurement of $\mathcal{R}(D)$ and $\mathcal{R}(D^*)$~\cite{Abdesselam:2019dgh}:
\begin{equation}
{\cal R}(D)^\text{Belle,2019}     = 0.307 \pm 0.037 \pm 0.016  \quad  \text{ and }   \quad
{\cal R}({D^*})^\text{Belle,2019} = 0.283 \pm 0.018 \pm 0.014\,,
\end{equation}
with a correlation of $-53\%$, and the combined results compatible with SM at the $1.2\sigma$ level. Including these new measurements in the global average leads to \cite{Amhis:2016xyh}
\begin{eqnarray}
{\cal R}(D)^{\text{avg,new}}   &= 0.340\pm 0.027 \pm0.013 \quad  \text{  and  }  \quad
{\cal R}(D^*)^{\text{avg,new}} &= 0.299\pm 0.013 \pm0.008\,,
\end{eqnarray}
and the significance of the anomaly amounts to  $3.1\sigma$ relative to the above SM prediction. In addition, LHCb also reported a value of the ratio as \cite{Aaij:2017tyk}
\begin{eqnarray}
{\cal R}(J/\psi)=\frac{\mathcal{B}(B_c\to J/\psi \tau\bar \nu)}{\mathcal{B}(B_c\to J/\psi \ell\bar \nu)}=0.71\pm 0.17\pm0.18.
\end{eqnarray}
This result deviates $2\sigma$ away from the SM predictions, which lies in the range of 0.23 to 0.28, where the uncertainties arises from the choice of modeling approach for the $B_c \to J/\psi$ form factors \cite{Wen-Fei:2013uea, Hu:2019qcn, Watanabe:2017mip, Tran:2018kuv, Bhattacharya:2018kig}.

Although these deviations are perhaps from the uncertainties of hadronic transition form factors in SM, they might imply that the lepton flavour universality (LFU) is violated, which is the hint of the existence of NP, because the LFU is one of the major characters of SM. Due to the fact that there is no any similar discrepancy in $K$ and $\pi$ semi-leptonic and purely leptonic decays, or in electroweak precision observables, most of us believe that LFU violation only appears in the third generation fermions. In this context, there are numerous works had been carried out based on model-independence approaches \cite{Huang:2018nnq, Bhattacharya:2018kig, Murgui:2019czp, Bernlochner:2017jka, Datta:2012qk, Bhattacharya:2016zcw, Duraisamy:2013kcw, Buttazzo:2017ixm, Alok:2017qsi, Bordone:2017anc, Jung:2018lfu, Azatov:2018knx, Hu:2018veh, Aebischer:2018iyb,Blanke:2018yud, Shi:2019gxi, Asadi:2019xrc, Aebischer:2019mlg,Ikeno:2019tkh,Dai:2019zur} or special models by introducing new particles such as a charged Higgs boson \cite{Tanaka:1994ay, Celis:2012dk, Celis:2016azn, Iguro:2017ysu, Fraser:2018aqj, Martinez:2018ynq}, leptoquarks \cite{Barbieri:2016las, Li:2016vvp, Bauer:2015knc, Fajfer:2015ycq, Freytsis:2015qca, Barbieri:2015yvd, Alonso:2015sja, Sakaki:2013bfa, Yan:2019hpm, Bordone:2017bld, Cornella:2019hct, Baker:2019sli, Angelescu:2018tyl, Fornal:2018dqn, Crivellin:2018yvo, Becirevic:2018afm, DiLuzio:2017vat, Assad:2017iib,Cai:2017wry, Becirevic:2016yqi,Crivellin:2019qnh, Hati:2019ufv}, or new vector bosons \cite{Megias:2017ove, He:2017bft, Matsuzaki:2017bpp, Babu:2018vrl, Asadi:2018wea, Greljo:2018ogz,Yang:2018pyq,Hu:2018lmk}.

If the ${\cal R}(D^{(*)})$ anomalies are due to NP, it is natural for us to ask whether its effects can be shown up in other $b \to c \tau \bar \nu_\tau$ transition decays, such as $B_c \to \eta_c \tau \bar \nu_\tau$ \cite{Huang:2018nnq, Bhattacharya:2018kig, Tran:2018kuv}, $B_s \to D_s^{(*)} \tau \bar \nu_\tau$ \cite{Harrison:2017fmw, Monahan:2018lzv, McLean:2019qcx} and $\Lambda_b \to \Lambda_c \tau \bar \nu_\tau$ \cite{Gutsche:2015mxa,Chen:2001zc, Detmold:2015aaa, Datta:2017aue, Detmold:2016pkz, Li:2016pdv, DiSalvo:2018ngq, Ray:2018hrx, Bernlochner:2018kxh, Bernlochner:2018bfn, Shivashankara:2015cta}. In this work, we shall study the $\Lambda_b \to \Lambda_c \tau \bar \nu_\tau$ comprehensively in model-dependence and independence manners, as well as $\Lambda_b \to \Lambda_c \ell \bar \nu_\ell$ ($\ell=e,\mu$). In this work, we will calculate the total rates, the differential decay distributions, the longitudinal polarizations of the final states $\Lambda_c$ and $\tau$, and the forward-backward asymmetries in the lepton-side, with and without contributions of NP. Of course, the correlations between $\mathcal{R}(D^{(*)})$ and $\mathcal{R}(\Lambda_b)$ will be given. Compared with previous studies, there are some improvements in the current work: (i) we calculate the complete amplitude including the contributions of scalar and tensor operators, and the latter was usually neglected in many previous studies; (ii) we present the relations between different conventions of from factors, which are nontrivial in the calculations ; (iii) the $\Lambda_b \to \Lambda_c$ form factors we adopted are the latest results from lattice QCD calculations; (iv) in discussing the effects of NP, we refit the each wilson coefficient individually within the latest experimental averaged results, and the form factors of $B\to D^{(*)}$ in fitting are from the heavy quark effective theory including next-to-leading power ${\cal O} (1/m_b)$ and next-to-leading order ${\cal O} (\alpha_s)$ corrections.

Our paper is organized as follows: In Sec.~\ref{sec:AnalyticalFormulaes}, the analytical formulae in this work are presented, such as the effective hamiltonian, the form factors and the helicity amplitudes. All definitions of physical observables are also given in this section. In Sec.~\ref{sec:Results}, we will present the numerical results and discussions. At last, we will draw conclusions in Sec.~\ref{sec:Conclusions}.
\section{Analytical Formulae} \label{sec:AnalyticalFormulaes}
\subsection{The effective hamiltonian}
Within the presence of NP, the effective Lagrangian can be extended by incorporating new operator basis that includes all possible four-fermion interactions. Due to the absence of experimental evidence of deviations from the SM in tree-level transitions involving light leptons such as the precision measurements of the ratio $\mathcal{B}(\tau \to \mu \nu_{\tau} \bar{\nu}_{\mu})/\mathcal{B}(\tau \to e \nu_{\tau} \bar{\nu}_{e}) = 0.9762\pm 0.0028$ \cite{Tanabashi:2018oca}, the new physics is generally supposed to appear in the third generation fermions. Therefore, without considering the effects from the right-handed neutrinos, the effective Lagrangian of $b \to c \tau \bar\nu_\tau$ can be written as
\begin{equation}
{\cal L}_{eff} =- \frac{4G_F}{\sqrt2} V_{cb}\Big[ (1 + C_{VL}){\cal O}_{VL} + C_{VR}{\cal O}_{VR} + C_{SL}{\cal O}_{SL} + C_{SR}{\cal O}_{SR} + C_T{\cal O}_T \Big] + \text{h.c.} \,,\label{eq:lag}
\end{equation}
where $G_F$ is the Fermi constant and $V_{cb}$ is the CKM matrix element. The four-fermion operators can be defined as
\begin{eqnarray}
&{\cal O}_{VL}  = (\cbar\gamma^\mu P_L b)(\bar{\tau} \gamma_\mu P_L\nu_\tau) \,,  \,\,\,
 {\cal O}_{VR}  =  (\cbar\gamma^\mu P_R b)(\bar{\tau} \gamma_\mu P_L\nu_\tau) \,, \nonumber \\
& {\cal O}_{SL} =  (\cbar  P_L b)(\bar{\tau}   P_L\nu_\tau) \,, \,\,\,
  {\cal O}_{SR}  = (\cbar P_R b)(\bar{\tau} P_L\nu_\tau)\,, \nonumber \\
& {\cal O}_T =  (\cbar\sigma^{\mu\nu}P_L b)(\bar{\tau} \sigma_{\mu\nu}P_{L} \nu_\tau) \,,\label{eq:operators}
\end{eqnarray}
where $P_{L,R}=(1\mp \gamma_5)/2$ and $C_i$ ($i=VL$, $VR$, $SL$, $SR$, and $T$) are the corresponding wilson coefficients at the scale $\mu= m_b$, with $C_i=0$ in SM. It should be stressed that recent studies \cite{Cirigliano:2009wk, Alonso:2014csa, Cata:2015lta} based on the effective field theory of SM (SMEFT) \cite{Buchmuller:1985jz,Grzadkowski:2010es} showed that the operator ${\cal O}_{VR}$ does not contribute to LFU violation at leading order, however we here include this operator for completeness.

\subsection{Form factors} \label{sec:form factors}
In the calculation, the most important inputs are the hadronic transition form factors. The hadronic matrix elements of the vector and axial-vector currents between the two spin-half baryons $\Lambda_b$ and $\Lambda_c$ can be parameterized in terms of three form factors, respectively, as~\cite{Gutsche:2015mxa}
\begin{align}
\langle \Lambda_c,\lambda_2|\bar c\gamma_\mu b|\Lambda_b,\lambda_1\rangle=&\bar u_2(p_2,\lambda_2)\left[F_1^V(q^2)\gamma_\mu-\frac{F_2^V(q^2)}{M_{\Lambda_b}}i\sigma_{\mu\nu}q^\nu+\frac{F_3^V(q^2)}
{M_{\Lambda_b}}q_\mu\right]u_1(p_1,\lambda_1)\,,\label{eq:vector ffs}\\[0.2cm]
\langle \Lambda_c,\lambda_2|\bar c\gamma_\mu\gamma_5 b|\Lambda_b,\lambda_1\rangle=&\bar u_2(p_2,\lambda_2)\left[F_1^A(q^2)\gamma_\mu-\frac{F_2^A(q^2)}{M_{\Lambda_b}}i\sigma_{\mu\nu}q^\nu+\frac{F_3^A(q^2)}
{M_{\Lambda_b}}q_\mu\right]\gamma_5u_1(p_1,\lambda_1)\,,\label{eq:axial-vector ffs}
\end{align}
where $\sigma_{\mu\nu}=\frac{i}{2}(\gamma_\mu\gamma_\nu-\gamma_\nu\gamma_\mu)$, $q=p_1-p_2$. $\lambda_i=\pm\frac{1}{2}\,(i=1,2)$ denote the helicities of the $\Lambda_b$ and $\Lambda_c$ baryons, respectively. Using the equations of motion, the ones of the scalar and pseudo-scalar currents can be obtained as
\begin{align}
\langle \Lambda_c,\lambda_2|\bar cb|\Lambda_b,\lambda_1\rangle=&\frac{1}{m_b-m_c}\bar u_2(p_2,\lambda_2)\left[F_1^V(q^2)M_-+\frac{F_3^V(q^2)}
{M_{\Lambda_b}}q^2\right]u_1(p_1,\lambda_1)\,,\label{eq:scalar ffs}\\[0.2cm]
\langle \Lambda_c,\lambda_2|\bar c\gamma_5b|\Lambda_b,\lambda_1\rangle=&\frac{1}{m_b+m_c}\bar u_2(p_2,\lambda_2)\left[F_1^A(q^2)M_+-\frac{F_3^A(q^2)}
{M_{\Lambda_b}}q^2\right]\gamma_5u_1(p_1,\lambda_1)\,,\label{eq:pseudo-scalar ffs}
\end{align}
and $m_b$ and $m_c$ are the current quark masses evaluated at the scale $\mu\sim m_b$, and $M_{\pm}=M_{\Lambda_b}\pm M_{\Lambda_c}$. Similarly, the hadronic matrix elements of the tensor and pseudo-tensor currents between the $\Lambda_b$ and $\Lambda_c$ baryons can be generally parameterized into four form factors as \cite{Chen:2001zc}
\begin{align}
\langle \Lambda_c,\lambda_2|\bar ci\sigma_{\mu\nu}b|\Lambda_b,\lambda_1\rangle=&\bar u_2(p_2,\lambda_2)
\Big[F_T i\sigma_{\mu\nu} + F_T^V (\gamma_\mu q_\nu-\gamma_\nu q_\mu)+ F_T^P (\gamma_\mu P_\nu-\gamma_\nu P_\mu)\nonumber\\
&+ F_T^S (P_\mu q_\nu-P_\nu q_\mu)\Big] u_1(p_1,\lambda_1)\,,\label{eq:tensor ffs}\\
\langle \Lambda_c,\lambda_2|\bar ci\sigma_{\mu\nu}\gamma_5b|\Lambda_b,\lambda_1\rangle=&\bar u_2(p_2,\lambda_2)
\Big[G_T i\sigma_{\mu\nu} + G_T^V (\gamma_\mu q_\nu-\gamma_\nu q_\mu) + G_T^P (\gamma_\mu P_\nu-\gamma_\nu P_\mu)\nonumber\\
 &+ G_T^S (P_\mu q_\nu-P_\nu q_\mu)\Big] \gamma_5 u_1(p_1,\lambda_1)\,,\label{eq:pseudo-tensor ffs}
\end{align}
where $P=p_1+p_2$. Noted that the terms $ F_T^P$ and $G_T^P$ had been missed in some literatures, such as in ref.\cite{Chen:2001zc}. Furthermore, the $F_T^i$ and $G_T^i$ are related by using the identity $\sigma^{\mu \nu}\gamma_{5}=-\frac{i}{2}\epsilon^{\mu \nu \alpha \beta}\sigma_{\alpha \beta}$, and the relations are given as
\begin{eqnarray}
&G_{T}=F_{T}-F^{V}_{T}M_+-F^{P}_{T}M_-
-F^{S}_{T}Q_+\,,\nonumber\\[0.2cm]
&G^{V}_{T}=F^{P}_{T}-F^{S}_{T}M_-,\,\,
G^{P}_{T}=F^{V}_{T}+F^{S}_{T}M_+,\,\,
G^{S}_{T}=-F^{S}_{T}\,.
\end{eqnarray}
where $Q_\pm=M_\pm^2-q^2$, which means that only four of them are linearly independent.

Alternatively, one can use the helicity-based definition of the $\Lambda_b\to\Lambda_c$ form factors \cite{Feldmann:2011xf}, then the matrix elements of the vector and axial vector currents can be written in terms of another six helicity form factors $f_+$, $f_\perp$, $f_0$, $g_+$, $g_\perp$, and $g_0$  as follows:
\begin{align}
\langle \Lambda_c,\lambda_2|\bar c\gamma_\mu b|\Lambda_b,\lambda_1\rangle&=\bar{u}_2(p_2,\lambda_2)\Big[ f_0 (q^2)M_-\frac{q^\mu}{q^2} +f_+ (q^2)\frac{M_+}{Q_+}(P^{\mu}-M_+M_-\frac{q^\mu}{q^2})\nonumber\\
& +f_\perp (q^2)(\gamma^\mu - \frac{2M_{\Lambda_c}}{Q_+}p_1^{\mu} - \frac{2M_{\Lambda_b}}{Q_+}p_2^{\mu})\Big]\bar{u}_1(p_1,\lambda_1), \label{eq:VFF} \\
\langle \Lambda_c,\lambda_2|\bar{c}\gamma^\mu \gamma_5 b|\Lambda_b,\lambda_1\rangle&=-\bar{u}_2(p_2,\lambda_2)\gamma_5\Big[ g_0 (q^2)M_+\frac{q^\mu}{q^2}+g_+ (q^2)\frac{M_-}{Q_-}(P^{\mu}-M_+M_-\frac{q^\mu}{q^2}) \nonumber\\
& +g_\perp (q^2)(\gamma^\mu + \frac{2M_{\Lambda_c}}{Q_-}p_1^{\mu} - \frac{2M_{\Lambda_b}}{Q_-}p_2^{\mu})\Big]\bar{u}_1(p_1,\lambda_1). \label{eq:AFF}
\end{align}
Again, within the equation of motion, the matrix elements of the scalar and pseudoscalar currents can be obtained as:
\begin{eqnarray}
\langle \Lambda_c,\lambda_2|\bar{c} b|\Lambda_b,\lambda_1\rangle &=& f_0(q^2)  \frac{M_-}{m_b-m_c} \bar{u}_2(p_2,\lambda_2)\bar{u}_1(p_1,\lambda_1), \\
\langle \Lambda_c,\lambda_2|\bar{c}\gamma_5 b|\Lambda_b,\lambda_1\rangle &=& g_0(q^2)  \frac{M_+}{m_b+m_c} \bar{u}_2(p_2,\lambda_2)\gamma_5 \bar{u}_1(p_1,\lambda_1).
\end{eqnarray}
Similarly, the matrix elements of the tensor current can also be written within the four helicity form factors $h_+$, $h_\perp$, $\widetilde{h}_+$, $\widetilde{h}_\perp$ \cite{Datta:2017aue},
\begin{align}
\langle \Lambda_c,\lambda_2|\bar ci\sigma_{\mu\nu}b|\Lambda_b,\lambda_1\rangle&=\bar u_2(p_2,\lambda_2)\Big[2h_+(q^2)\frac{p_1^\mu p_2^{ \nu}-p_1^\nu p_2^{\mu}}{Q_+} \nonumber\\
&+h_\perp (q^2)\Big(\frac{M_+}{q^2}(q^\mu \gamma^\nu -q^\nu \gamma^\mu)-2(\frac{1}{q^2}+\frac{1}{Q_+})(p_1^\mu p_2^{\nu}-p_1^\nu p_2^{\mu}) \Big) \nonumber\\
&+\widetilde{h}_+ (q^2)\Big(i\sigma^{\mu \nu}-\frac{2}{Q_-}(M_{\Lambda_b}(p_2^{\mu}\gamma^\nu -p_2^{\nu}\gamma^\mu)
-M_{\Lambda_c}(p_1^\mu \gamma^\nu -p_1^\nu \gamma^\mu)+p_1^\mu p_2^{\nu}-p_1^\nu p_2^{\mu}) \Big) \nonumber\\
&+\widetilde{h}_\perp(q^2) \frac{M_-}{q^2 Q_-}\Big((M_+M_--q^2)(\gamma^\mu p_1^\nu - \gamma^\nu p_1^\mu)\nonumber\\
&-(M_+M_-+q^2)(\gamma^\mu p_2^{\nu}-\gamma^\nu p_2^{\mu})+2 M_-(p_1^\mu p_2^{\nu}-p_1^\nu p_2^{\mu}) \Big)
\Big]\bar u_1(p_1,\lambda_1). \label{eq:TFF}
\end{align}
Of course, these two definitions are physically equivalent, and the relations between two groups of form factors are given as
\begin{align}
F^V_1(q^2)&=\frac{f_+(q^2)M_+^{2}-f_{\perp}(q^2)q^2}{Q_+}\,, \nonumber \\
F^V_2(q^2)&=\frac{(f_{\perp}(q^2)-f_+(q^2))M_{\Lambda_b}M_+}{Q_+}\,, \nonumber \\
F^V_3(q^2)&=\frac{M_-M_{\Lambda_b}\big[\big(f_0(q^2)-f_+(q^2)\big)M_+^{2}-\big(f_0(q^2)-f_{\perp}(q^2)\big)q^2\big]}{q^2 Q_+}\,, \nonumber \\
F^A_1(q^2)&=\frac{g_+(q^2)M_-^{2}-g_{\perp}(q^2)q^2}{Q_-}\,, \nonumber \\
F^A_2(q^2)&=\frac{(g_{\perp}(q^2)-g_+(q^2))M_{\Lambda_b}M_-}{Q_-}\,, \nonumber \\
F^A_3(q^2)&=\frac{M_+M_{\Lambda_b}\big[\big(g_+(q^2)-g_0(q^2)\big)M_-^{2}+\big(g_0(q^2)-g_{\perp}(q^2)\big)q^2\big]}{q^2 Q_-}\,, \nonumber \\
F_{T}(q^2)&=\widetilde{h}_{+}(q^2)\,,\nonumber\\
F_T^V(q^2)&=-\widetilde{h}_{+}(q^2)\frac{M_{+}}{Q_{-}}+\widetilde{h}_{\bot}(q^2)\frac{M^{2}_{-}M_{+}}{q^{2}Q_{-}}
-h_{\bot}(q^2)\frac{M_{+}}{q^{2}}\,,\nonumber\\
F_T^P(q^2)&=\widetilde{h}_{+}(q^2)\frac{M_{-}}{Q_{-}}-\widetilde{h}_{\bot}(q^2)\frac{M_{-}}{Q_{-}}\,,\nonumber\\
F_T^S(q^2)&=\widetilde{h}_{+}(q^2)\frac{1}{Q_{-}}-h_{+}(q^2)\frac{1}{Q_{+}}-\widetilde{h}_{\bot}(q^2)
\frac{M^{2}_{-}}{q^{2}Q_{+}}
+h_{\bot}(q^2)\frac{M^{2}_{+}}{q^{2}Q_{-}}\,.
\end{align}
For the various helicity form factors we have used the formulae from the latest  lattice QCD calculations \cite{Detmold:2015aaa, Datta:2017aue, Detmold:2016pkz}, and each form factor can be written as
\begin{eqnarray}
&&f(q^2) = \frac{1}{1-q^2/(m_{\rm pole}^f)^2}\,\Big[a_0^f + a_1^f\,z(q^2)\Big]\,,
\end{eqnarray}
where $m_{\rm pole}^f$ is pole mass, and $f$ represents $f_{+,{\perp},\,0}$, $g_{+,{\perp},\,0}$, $h_{+,\perp}$ and  $\widetilde{h}_{+,\perp}$, respectively. The explicit numerical values of $m_{\rm pole}^f$, $a_0^f$, and $a_1^f$ can be found in refs.\cite{Detmold:2015aaa, Datta:2017aue, Detmold:2016pkz}, and we will not list them in the current work. Besides, the definition of the expansion parameter $z$ is
\begin{eqnarray}
&&z(q^2) = \frac{\sqrt{t_+ - q^2} - \sqrt{t_+ - t_0}}{\sqrt{t_+ - q^2} + \sqrt{t_+ - t_0}}\,,
\end{eqnarray}
where $t_+ = (M_{\Lambda_b} + M_{\Lambda_c})^2$ and $t_0 = (M_{\Lambda_b} - M_{\Lambda_c})^2$, respectively.

\subsection{The helicity amplitude}
\begin{figure}[!htb]
\begin{center}
\includegraphics[width=8cm,height=5cm]{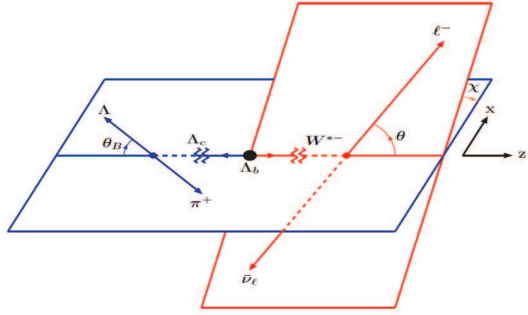}
\caption{Definition of the polar angles $\theta$, $\theta_B$ and the azimuthal angle $\chi$.}\label{fig:angles}
\end{center}
\end{figure}

In SM, the semi-leptonic decay $\Lambda_b\to\Lambda_c \ell^-\bar\nu_\ell$ is considered to be $\Lambda_b\to\Lambda_c W^{-}_{\rm off-shell}$, and the off-shell $W^{-}_{\rm off-shell}$ decays to $\ell^-\bar\nu_l$, subsequently. It is known to us that the $W^{-}_{\rm off-shell}$ has four helicities, namely $\lambda_W=\pm 1,0\,(J=1)$ and $\lambda_W=0\,(J=0)$, and only the off-shell $W^{-}_{\rm off-shell}$ has a timelike polarization, with $J=1,0$ denoting the two angular momenta of the rest frame $W^{-}_{\rm off-shell}$. In order to distinguish the two $\lambda_W=0$ states we adopt the notation $\lambda_W=0$ for $J=1$ and $\lambda_W=t$ for $J=0$. In the $\Lambda_b$-baryon rest frame, choosing the $z$--axis to be along the $W^{-}_{\rm off-shell}$ (see Fig.~\ref{fig:angles}), we can write the polarization of the $W^{-}_{\rm off-shell}$ as
\begin{eqnarray}
\epsilon^\mu(\pm)=\frac{1}{\sqrt{2}}(0,1,\mp i,0);\,\,\,\,
\epsilon^\mu(0)=-\frac{1}{\sqrt{q^2}}(q_z,0,0,q_0); \,\,\,\
\epsilon^\mu(t)=-\frac{q^\mu}{\sqrt{q^2}};
\end{eqnarray}
where $q^\mu$ is the four-momentum of the $W^{-}_{\rm off-shell}$. In this case, the polarization vectors of the $W^{-}_{\rm off-shell}$ satisfy the orthonormality and completeness relations:
\begin{eqnarray}
\epsilon^{*\mu}(m)\epsilon_\mu(n)=g_{mn},\,\,
\sum_{m,n}\epsilon^{*\mu}(m)\epsilon^\nu(n)g_{mn}=g^{\mu\nu}
\end{eqnarray}
with $g_{mn}={\rm diag}(+,-,-,-)$ for $m,n=t,\pm,0$. Because the current is not conserved in SM, so it consists of a superposition of a spin-1 and a spin-0 component where the $J^{P}$ content of the vector current $J^V_{\mu}$ and the axial vector current $J^A_{\mu}$ are $(0^{+},1^{-})$ and $(0^{-},1^{+})$, respectively. Therefore, we define helicity amplitudes as
\begin{equation}
H_{\lambda_2,\lambda_W}^{L}=H_{\lambda_2,\lambda_W}^V -H_{\lambda_2,\lambda_W}^A\,,\quad H_{\lambda_2,\lambda_W}^{V(A)}=\epsilon^{\dag\mu}(\lambda_W)\langle \Lambda_c,\lambda_2|\bar c\gamma_\mu(\gamma_\mu\gamma_5) b|\Lambda_b,\lambda_1\rangle\,.
\end{equation}
Due to the conservation of angular momentum,  $\lambda_{1}=-\lambda_{2}+\lambda_{W}$ is satisfied.
Wth above definitions, we preform the helicity amplitudes and have (see e.g.  Refs.~\cite{Gutsche:2013pp, Gutsche:2013oea, Gutsche:2014zna})
\begin{eqnarray}
H_{+\frac12 t}^{V/A}&=&\frac{\sqrt{Q_\pm}}{\sqrt{q^2}}
  \bigg( M_\mp F_1^{V/A}\pm \frac{q^2}{M_{\Lambda_b}} F_3^{V/A}\bigg),
 \nonumber\\
H_{+\frac12 +1}^{V/A}&=&\sqrt{2Q_\mp}
  \bigg(F_1^{V/A}\pm \frac{M_\pm}{M_{\Lambda_b}}F_2^{V/A}\bigg),
 \nonumber\\
H_{+\frac12 0}^{V/A}&=&\frac{\sqrt{Q_\mp}}{\sqrt{q^2}}
  \bigg(M_\pm F_1^{V/A}\pm \frac{q^2}{M_{\Lambda_b}} F_2^{V/A}\bigg)\,.
\label{eq:hel_inv}
\end{eqnarray}
From parity or from an explicit calculation, we have
\begin{eqnarray}
H_{-\lambda_2,-\lambda_W}^V  = H_{\lambda_2,\lambda_W}^V,
\qquad
H_{-\lambda_2,-\lambda_W}^A = -H_{\lambda_2,\lambda_W}^A.
\end{eqnarray}
If the right-handed current exists, we also have
\begin{equation}
H_{\lambda_2,\lambda_W}^{R}=H_{\lambda_2,\lambda_W}^V+H_{\lambda_2,\lambda_W}^A\,.
\end{equation}
For the $(S\mp P)$-type current, the corresponding helicity amplitudes are given by~\cite{Shivashankara:2015cta}
\begin{align}\label{eq:scalar1 amplitudes}
H^{SPL}_{\pm\frac{1}{2},0}=&\frac{\sqrt{Q_+}}{m_b-m_c}\left(F_1^VM_-+F_3^V\frac{q^2}{M_{\Lambda_b}}\right)\pm\frac{\sqrt{Q_-}}
{m_b+m_c}\left(F_1^AM_+-F_3^A\frac{q^2}{M_1}\right)\,,\\
H^{SPR}_{\pm\frac{1}{2},0}=&\frac{\sqrt{Q_+}}{m_b-m_c}\left(F_1^VM_-+F_3^V\frac{q^2}{M_{\Lambda_b}}\right)\mp\frac{\sqrt{Q_-}}
{m_b+m_c}\left(F_1^AM_+-F_3^A\frac{q^2}{M_1}\right)\,.
\end{align}
Also, one could define the hadronic helicity amplitudes of the (pseudo-)tensor-type current as
\begin{equation}\label{eq:tensor definitions}
H_{\lambda_2,\lambda_W,\lambda_{W^\prime}}^{T}=\epsilon^{\dag\mu}(\lambda_W)\epsilon^{\dag\nu}
(\lambda_W^\prime)\langle \Lambda_c,\lambda_2|\bar c\,i\sigma_{\mu\nu}(1-\gamma_5)b|\Lambda_b,\lambda_1\rangle\,,
\end{equation}
and their explicit expressions are given by
\begin{align}\label{eq:tensor amplitudes}
H_{\frac{1}{2},+,0}^T=&-\sqrt{\frac{2}{q^2}}\left(F_T\sqrt{Q_+}M_--F^{P}_{T}\sqrt{Q_+}Q_-+G_T\sqrt{Q_-}M_+
+G^{P}_{T}\sqrt{Q_-}Q_+\right)\,,\nonumber\\[0.2cm]
H_{\frac{1}{2},+,-}^T=&-F_T\sqrt{Q_+}-G_T\sqrt{Q_-}\,,\nonumber\\
H_{\frac{1}{2},+,t}^T=&-\sqrt{\frac{2}{q^2}}\left(F_T\sqrt{Q_-}M_+-F^{P}_{T}\sqrt{Q_-}M_{+}M_{-}+G_T\sqrt{Q_+}M_-
+G^{P}_{T}\sqrt{Q_+}M_{+}M_{-}\right)\nonumber\\[0.2cm]
&\hspace{0.0cm}+\sqrt{2q^2}\left(F_T^V\sqrt{Q_-}-G_T^V\sqrt{Q_+}\right)\,,\nonumber\\
H_{\frac{1}{2},0,t}^T=&-F_T\sqrt{Q_-}-G_T\sqrt{Q_+}+F^{P}_{T}\sqrt{Q_-}M_{-}-G^{P}_{T}\sqrt{Q_+}M_{+}\nonumber\\
&+F_T^V\sqrt{Q_-}M_+-G_T^V\sqrt{Q_+}M_-+F_T^S\sqrt{Q_-}Q_++G_T^S\sqrt{Q_+}Q_-\,,\nonumber\\
H_{-\frac{1}{2},+,-}^T=&F_T\sqrt{Q_+}-G_T\sqrt{Q_-}\,,\nonumber\\[0.2cm]
H_{-\frac{1}{2},0,-}^T=&\sqrt{\frac{2}{q^2}}\left(F_T\sqrt{Q_+}M_--F^{P}_{T}\sqrt{Q_+}Q_-
-G_T\sqrt{Q_-}M_+-G^{P}_{T}\sqrt{Q_-}Q_+\right)\,,\nonumber\\[0.2cm]
H_{-\frac{1}{2},0,t}^T=&-F_T\sqrt{Q_-}+G_T\sqrt{Q_+}+F^{P}_{T}\sqrt{Q_-}M_{-}+G^{P}_{T}\sqrt{Q_+}M_{+}\nonumber\\
&+F_T^V\sqrt{Q_-}M_++G_T^V\sqrt{Q_+}M_-+F_T^S\sqrt{Q_-}Q_+-G_T^S\sqrt{Q_+}Q_-\,,\nonumber\\[0.2cm]
H_{-\frac{1}{2},-,t}^T=&-\sqrt{\frac{2}{q^2}}\left(F_T\sqrt{Q_-}M_+-F^{P}_{T}\sqrt{Q_-}M_{+}M_{-}
-G_T\sqrt{Q_+}M_--G^{P}_{T}\sqrt{Q_+}M_{+}M_{-}\right)\nonumber\\[0.2cm]
&+\sqrt{2q^2}\left(F_T^V\sqrt{Q_-}+G_T^V\sqrt{Q_+}\right)\,.
\end{align}

\subsection{Observable }
With the above hadronic helicity amplitudes, we then write down the two-fold differential angular decay distribution as
\begin{equation}\label{eq:differential angular}
\frac{{\rm d}^2\Gamma(\Lambda_b\to\Lambda_c\ell\bar\nu_\ell)}{{\rm d}q^2\,{\rm d}\cos\theta}=N\left[A_1+\frac{m_\ell^2}{q^2}(A_{2}^{VL}+A_{2}^{VR}+A_{2}^{LR}+A_{2}^{T})+2A_3+
\frac{4m_\ell}{\sqrt{q^2}}(A_{4}^{VS}+A_{4}^{LT}+A_{4}^{RT})+A_5\right]\,,
\end{equation}
with
\begin{align}\label{eq:norm}
N=&\frac{G_F^2|V_{cb}|^2q^2|\vec{\mathbf p}_2|}{512\pi^3M_{\Lambda_b}^2}\left(1-\frac{m_\ell^2}{q^2}\right)^2\,.
\end{align}
In above equation, $|\vec{\mathbf p}_2|=\sqrt{Q_{+}Q_{-}}/(2M_{\Lambda_b})$ denotes the $\Lambda_c$ momentum in the $\Lambda_b$ rest frame. $\theta$ is the polar angle of the lepton, the definition of which can be seen in Fig.\ref{fig:angles}. The auxiliary functions $A_i$ are given as
\begin{align}\label{eq:coefficients1}
A_1=&|1+C_{VL}|^2\left[2\sin^2\theta\big(|H_{\frac{1}{2},0}^{L}|^2+|H_{-\frac{1}{2},0}^{L}|^2\big)+(1-\cos\theta)^2
|H_{\frac{1}{2},+}^{L}|^2+(1+\cos\theta)^2|H_{-\frac{1}{2},-}^{L}|^2\right]\nonumber\\[0.1cm]+&
|C_{VR}|^2\left[2\sin^2\theta\big(|H_{\frac{1}{2},0}^{R}|^2+|H_{-\frac{1}{2},0}^{R}|^2\big)+(1-\cos\theta)^2
|H_{\frac{1}{2},+}^{R}|^2+(1+\cos\theta)^2|H_{-\frac{1}{2},-}^{R}|^2\right]\,,\\[0.3cm]
A_2^{VL}=&|1+C_{VL}|^2\Big[2\cos^2\theta\big(|H_{\frac{1}{2},0}^{L}|^2+|H_{-\frac{1}{2},0}^{L}|^2\big)+\sin^2\theta
\big(|H_{\frac{1}{2},+}^{L}|^2+|H_{-\frac{1}{2},-}^{L}|^2\big)
+2\big(|H_{\frac{1}{2},t}^{L}|^2+|H_{-\frac{1}{2},t}^{L}|^2\big)\nonumber\\[0.1cm]
&\,\,\,\,-4\cos\theta\,\big(H_{\frac{1}{2},0}^{L}H_{\frac{1}{2},t}^{L}+H_{-\frac{1}{2},0}^{L} H_{-\frac{1}{2},t}^{L}\big)\Big] \,,\\[0.3cm]
A_2^{VR}=&|C_{VR}|^2\Big[2\cos^2\theta\big(|H_{\frac{1}{2},0}^{R}|^2+|H_{-\frac{1}{2},0}^{R}|^2\big)+\sin^2\theta
\big(|H_{\frac{1}{2},+}^{R}|^2+|H_{-\frac{1}{2},-}^{R}|^2\big)
+2\big(|H_{\frac{1}{2},t}^{R}|^2+|H_{-\frac{1}{2},t}^{R}|^2\big)\nonumber\\[0.1cm]
&\,\,\,\,-4\cos\theta\,\big(H_{\frac{1}{2},0}^{R}H_{\frac{1}{2},t}^{R}+H_{-\frac{1}{2},0}^{R} H_{-\frac{1}{2},t}^{R}\big)\Big] \,,\\[0.3cm]
A_{2}^{LR}=&2\,{\rm Re}\big[(1+C_{VL})C_{VR}^{*}\big]\Big[\sin^2\theta\big(H_{-\frac{1}{2},-}^{L}H_{-\frac{1}{2},-}^{R}
+H_{\frac{1}{2},+}^{L}H_{\frac{1}{2},+}^{R}\big)
+2\cos^{2}\theta\,\big(H_{-\frac{1}{2},0}^{L}H_{-\frac{1}{2},0}^{R}+H_{\frac{1}{2},0}^{L}H_{\frac{1}{2},0}^{R}\big)
\nonumber\\[0.1cm]
&-2\cos\theta\,\big(H_{-\frac{1}{2},t}^{L}H_{-\frac{1}{2},0}^{R}+H_{-\frac{1}{2},0}^{L}H_{-\frac{1}{2},t}^{R}
+H_{\frac{1}{2},t}^{L}H_{\frac{1}{2},0}^{R}+H_{\frac{1}{2},0}^{L}H_{\frac{1}{2},t}^{R}\big)
+2\,\big(H_{-\frac{1}{2},t}^{L}H_{-\frac{1}{2},t}^{R}
+H_{\frac{1}{2},t}^{L}H_{\frac{1}{2},t}^{R}\big)
\Big] \,,\\[0.3cm]
A_2^T=&4\,|C_T|^2\Big[2\sin^2\theta\big(|H_{\frac{1}{2},+,-}^{T}+H_{\frac{1}{2},0,t}^{T}|^{2}+
|H_{-\frac{1}{2},+,-}^{T}+H_{-\frac{1}{2},0,t}^{T}|^{2}\big)\nonumber\\[0.1cm]
&\hspace{1.1cm}+(1+\cos\theta)^2|H_{-\frac{1}{2},0,-}^{T}+H_{-\frac{1}{2},-,t}^{T}|^{2}
+(1-\cos\theta)^2|H_{\frac{1}{2},+,0}^{T}+H_{\frac{1}{2},+,t}^{T}|^{2}\Big]\,,\\[0.3cm]
A_3=&2|C_T|^2\Big[2\cos^2\theta\big(|H_{\frac{1}{2},+,-}^{T}+H_{\frac{1}{2},0,t}^{T}|^{2}
+|H_{-\frac{1}{2},+,-}^{T}+H_{-\frac{1}{2},0,t}^{T}|^{2}\big)\nonumber\\[0.1cm]
&\hspace{1.1cm}+\sin^2\theta\big(
|H_{\frac{1}{2},+,0}^{T}\!+\!H_{\frac{1}{2},+,t}^{T}|^{2}+|H_{-\frac{1}{2},0,-}^{T}+H_{-\frac{1}{2},-,t}^{T}|^{2}
\big)\Big]\nonumber\\[0.1cm]
&\hspace{0.4cm}+|C_{SL}|^2\big(|H^{SPL}_{\frac{1}{2},0}|^{2}+|H^{SPL}_{-\frac{1}{2},0}|^{2}\big)
+|C_{SR}|^2\big(|H^{SPR}_{\frac{1}{2},0}|^{2}+|H^{SPR}_{-\frac{1}{2},0}|^{2}\big)\,,\\[0.3cm]
A_4^{VS}=&{\rm Re} \big[(1+C_{VL})C_{SL}^{*}\big]\big[-\cos\theta\,\big(H_{\frac{1}{2},0}^{L}H^{SPL}_{\frac{1}{2},0}
+H_{-\frac{1}{2},0}^{L}H^{SPL}_{-\frac{1}{2},0}\big)
+H_{\frac{1}{2},t}^{L}H^{SPL}_{\frac{1}{2},0}+H_{-\frac{1}{2},t}^{L}H^{SPL}_{-\frac{1}{2},0}\big]\nonumber\\[0.1cm]
&+{\rm Re}\big[(1+C_{VL})C_{SR}^{*}\big]\big[-\cos\theta\,\big(H_{-\frac{1}{2},0}^{L}H^{SPR}_{-\frac{1}{2},0}
+H_{\frac{1}{2},0}^{L}H^{SPR}_{\frac{1}{2},0}\big)
+H_{\frac{1}{2},t}^{L}H^{SPR}_{\frac{1}{2},0}+H_{-\frac{1}{2},t}^{L}H^{SPR}_{-\frac{1}{2},0}\big]\nonumber\\[0.1cm]
&+{\rm Re}\big[C_{VR}C_{SL}^{*}\big]\big[-\cos\theta\,\big(H_{\frac{1}{2},0}^{R}H^{SPL}_{\frac{1}{2},0}
+H_{-\frac{1}{2},0}^{R}H^{SPL}_{-\frac{1}{2},0}\big)
+H_{\frac{1}{2},t}^{R}H^{SPL}_{\frac{1}{2},0}+H_{-\frac{1}{2},t}^{R}H^{SPL}_{-\frac{1}{2},0}\big]\nonumber\\[0.1cm]
&+{\rm Re}\big[C_{VR}C_{SR}^{*}\big]\big[-\cos\theta\,\big(H_{\frac{1}{2},0}^{R}H^{SPR}_{\frac{1}{2},0}
+H_{-\frac{1}{2},0}^{R}H^{SPR}_{-\frac{1}{2},0}\big]
+H_{\frac{1}{2},t}^{R}H^{SPR}_{\frac{1}{2},0}+H_{-\frac{1}{2},t}^{R}H^{SPR}_{-\frac{1}{2},0}\big]\\[0.1cm]
A_4^{LT}=&-2{\rm Re}\big[(1+C_{VL})C_{T}^{*}\big]\Big\{(1+\cos\theta)\,H_{-\frac{1}{2},-}^{L}\big(H^{T}_{-\frac{1}{2},-,t}
+H^{T}_{-\frac{1}{2},0,-}\big)+(1-\cos\theta)\,\big(H_{\frac{1}{2},+}^{L}H^{T}_{\frac{1}{2},+,t}\big)\nonumber\\[0.1cm]
&\hspace{3.1 cm}
+\,H_{-\frac{1}{2},0}^{L}\big(H^{T}_{-\frac{1}{2},0,t}+H^{T}_{-\frac{1}{2},+,-}\big)
+H_{\frac{1}{2},0}^{L}\big(H^{T}_{\frac{1}{2},0,t}+H^{T}_{\frac{1}{2},+,-}\big)
+H_{\frac{1}{2},+}^{L}H^{T}_{\frac{1}{2},+,0}\nonumber\\[0.1cm]
&\hspace{2.0cm}
-\cos\theta\,\big[H_{-\frac{1}{2},t}^{L}\big(H^{T}_{-\frac{1}{2},0,t}+H^{T}_{-\frac{1}{2},+,-}\big)
+H_{\frac{1}{2},t}^{L}\big(H^{T}_{\frac{1}{2},0,t}+H^{T}_{\frac{1}{2},+,-}\big)
+H_{\frac{1}{2},+}^{L}H^{T}_{\frac{1}{2},+,0}\big]
\Big\}\,,\\[0.3cm]
A_4^{RT}=&-2{\rm Re}\big(C_{VR}C_{T}^{*}\big)\Big\{(1+\cos\theta)\,H_{-\frac{1}{2},-}^{R}\big(H^{T}_{-\frac{1}{2},-,t}
+H^{T}_{-\frac{1}{2},0,-}\big)
+(1-\cos\theta)\,\big(H_{\frac{1}{2},+}^{R}H^{T}_{\frac{1}{2},+,t}\big)\nonumber\\[0.1cm]
&\hspace{3.1 cm}
+\,H_{-\frac{1}{2},0}^{R}\big(H^{T}_{-\frac{1}{2},0,t}+H^{T}_{-\frac{1}{2},+,-}\big)
+H_{\frac{1}{2},0}^{R}\big(H^{T}_{\frac{1}{2},0,t}+H^{T}_{\frac{1}{2},+,-}\big)
+H_{\frac{1}{2},+}^{R}H^{T}_{\frac{1}{2},+,0}\nonumber\\[0.1cm]
&\hspace{2.0cm}
-\cos\theta\,\big[H_{-\frac{1}{2},t}^{R}\big(H^{T}_{-\frac{1}{2},0,t}+H^{T}_{-\frac{1}{2},+,-}\big)
+H_{\frac{1}{2},t}^{R}\big(H^{T}_{\frac{1}{2},0,t}+H^{T}_{\frac{1}{2},+,-}\big)
+H_{\frac{1}{2},+}^{R}H^{T}_{\frac{1}{2},+,0}\big]\Big\}\,,\\[0.3cm]
A_5=&8\cos\theta\,{\rm Re}\big(C_{SL}C_T^{*}\big)\Big[H^{SPL}_{\frac{1}{2},0}\big(H^T_{\frac{1}{2},+,-}+H^T_{\frac{1}{2},0,t}\big)
+H^{SPL}_{-\frac{1}{2},0}\big(H^T_{-\frac{1}{2},+,-}
+H^T_{-\frac{1}{2},0,t}\big)\Big]\nonumber\\&+8\cos\theta\,{\rm Re}\big(C_{SR}C_T^{*}\big)\Big[H^{SPR}_{\frac{1}{2},0}\big(H^T_{\frac{1}{2},+,-}+
H^T_{\frac{1}{2},0,t}\big)+H^{SPR}_{-\frac{1}{2},0}\big(H^T_{-\frac{1}{2},+,-}
+H^T_{-\frac{1}{2},0,t}\big)\Big]\nonumber\\[0.1cm]
&+2{\rm Re}\big[(1+C_{VL})C_{VR}^{*}\big]\Big[(1+\cos\theta)^{2}\,H^{L}_{-\frac{1}{2},-}H^{R}_{-\frac{1}{2},-}
+(1-\cos\theta)^{2}\,H^{L}_{\frac{1}{2},+}H^{R}_{\frac{1}{2},+}
\nonumber\\[0.1cm]&\hspace{1.1cm}+
2\sin^{2}\theta\,\big(H^{L}_{-\frac{1}{2},0}H^{R}_{-\frac{1}{2},0}+H^{L}_{\frac{1}{2},0}H^{R}_{\frac{1}{2},0}\big)\Big]+4\,\,
{\rm Re}\big(C_{SL}C_{SR}^{*}\big)\Big(H^{SPL}_{-\frac{1}{2},0}H^{SPR}_{-\frac{1}{2},0}+
H^{SPL}_{\frac{1}{2},0}H^{SPR}_{\frac{1}{2},0}\Big).\label{eq:coefficients2}
\end{align}
When integrating out $\cos\theta$ in eq.~\eqref{eq:differential angular},  we then obtain the differential decay rate ${\rm d}\Gamma(\Lambda_b\to\Lambda_c\ell\bar\nu_{\ell})/{\rm d}q^2$ and the total branching fraction can be given as
\begin{equation}
\mathcal B(\Lambda_b\to\Lambda_c\ell\bar\nu_\ell)=\tau_{\Lambda_b}\int_{m_{\ell}^2}^{M_{-}^2} {\rm d}q^2 \frac{{\rm d}\Gamma(\Lambda_b\to\Lambda_c\tau\bar\nu_\tau)}{{\rm d}q^2}\,,
\end{equation}
$\tau_{\Lambda_b}$  denoting the lifetime of $\Lambda_b$ baryon. Similar to $\mathcal{R}(D^{(*)})$, one can define the differential and integrated ratios as
\begin{eqnarray}
 &&\mathcal{R}(\Lambda_c)(q^2)=\frac{{\rm d}\Gamma(\Lambda_b\to\Lambda_c\tau\bar\nu_\tau)/{\rm d}q^2}{{\rm d}\Gamma(\Lambda_b\to\Lambda_c\mu\bar\nu_\mu)/{\rm d}q^2}\,,\\
 &&\mathcal{R}(\Lambda_c)=\frac{\int_{m_{\tau}^2}^{M_{-}^2} {\rm d}q^2{\rm d}\Gamma(\Lambda_b\to\Lambda_c\tau\bar\nu_\tau)/{\rm d}q^2}{\int_{m_{\mu}^2}^{M_{-}^2} {\rm d}q^2{\rm d}\Gamma(\Lambda_b\to\Lambda_c\mu\bar\nu_\mu)/{\rm d}q^2}\,.
\end{eqnarray}
Also, the forward-backward asymmetry in the lepton-side is defined as
 \begin{equation}
  A_{\rm FB}(q^2) = \frac{\int_{0}^{1} {\rm d}\cos\theta({\rm d}^2\Gamma/{\rm d}q^2{\rm d}\cos\theta)-\int_{-1}^{0}{\rm d}\cos\theta({\rm d}^2\Gamma/{\rm d}q^2{\rm d} \cos\theta )}{\int_{0}^{1} {\rm d}\cos\theta({\rm d}^2\Gamma/{\rm d}q^2{\rm d}\cos\theta)+\int_{-1}^{0}{\rm d}\cos\theta({\rm d}^2\Gamma/{\rm d}q^2{\rm d} \cos\theta )}\,.
  \end{equation}
Besides, we can calculate the helicity-dependent differential decay rates, which are given as (normalized by the prefactor $N/2$) :
\begin{align}
\frac{{\rm d}\Gamma^{\lambda_{2}=1/2}}{{\rm d}q^2}=&\frac{m_\ell^2}{q^2}\Big[\frac{8}{3}|1+C_{VL}|^2\big(|H^{L}_{\frac{1}{2},+}|^2+|H^{L}_{\frac{1}{2},0}|^2+
3|H^{L}_{\frac{1}{2},t}|^2\big)+\frac{8}{3}|C_{VR}|^2\big(|H^{R}_{\frac{1}{2},0}|^2+|H^{R}_{\frac{1}{2},+}|^2
+3|H^{R}_{\frac{1}{2},t}|^{2}\big)\nonumber\\[0.1cm]
&\hspace{0.5cm}+\frac{64}{3}|C_T|^2\big(|H^{T}_{\frac{1}{2},+,-}+H^{T}_{\frac{1}{2},0,t}|^{2}
+|H^{T}_{\frac{1}{2},+,0}+H^{T}_{\frac{1}{2},+,t}|^{2}\big)\nonumber\\[0.1cm]
&\hspace{0.5cm}+\frac{16}{3}{\rm Re}[(1+C_{VL})C_{VR}^{*}]\big(H^{L}_{\frac{1}{2},0}H^{R}_{\frac{1}{2},0}+H^{L}_{\frac{1}{2},+}H^{R}_{\frac{1}{2},+}
+3H^{L}_{\frac{1}{2},t}H^{R}_{\frac{1}{2},t}\big)\Big]\nonumber\\[0.1cm]
+&\frac{16m_\ell}{\sqrt{q^2}}{\rm Re}\big[[(1+C_{VL})H^{L}_{\frac{1}{2},t}+C_{VR}H^{R}_{\frac{1}{2},t}](C_{SL}^{*}H^{SPL}_{\frac{1}{2},0}
+C_{SR}^{*}H^{SPR}_{\frac{1}{2},0})\nonumber\\[0.1cm]
&\hspace{1.5cm}-2C_{T}^{*}[(1+C_{VL})H^{L}_{\frac{1}{2},0}+C_{VR}H^{R}_{\frac{1}{2},0}](H^{T}_{\frac{1}{2},0,t}
+H^{T}_{\frac{1}{2},+,-})
\nonumber\\[0.1cm]
&\hspace{1.5cm}-2C_{T}^{*}[(1+C_{VL})H^{L}_{\frac{1}{2},+}+C_{VR}H^{R}_{\frac{1}{2},+}](H^{T}_{\frac{1}{2},+,0}
+H^{T}_{\frac{1}{2},+,t})\big]\nonumber\\[0.1cm]
+&\frac{16}{3}\Big[|1+C_{VL}|^{2}(|H^{L}_{\frac{1}{2},0}|^{2}+|H^{L}_{\frac{1}{2},+}|^{2})+|C_{VR}|^{2}
(|H^{R}_{\frac{1}{2},0}|^{2}+
|H^{R}_{\frac{1}{2},+}|^{2})\nonumber\\[0.1cm]
&\hspace{0.5cm}+\frac{3}{2}|(C_{SL}H^{SPL}_{\frac{1}{2},0}+C_{SR}H^{SPR}_{\frac{1}{2},0})|^{2}
+2|C_{T}|^{2}\big(|H^{T}_{\frac{1}{2},0,t}+H^{T}_{\frac{1}{2},+,-}|^{2}+|H^{T}_{\frac{1}{2},+,0}
+H^{T}_{\frac{1}{2},+,t}|^{2}\big)\nonumber\\[0.1cm]
&\hspace{0.5cm}+2{\rm Re}[(1+C_{VL})C_{VR}^{*}]\big(H^{L}_{\frac{1}{2},0}H^{R}_{\frac{1}{2},0}+H^{L}_{\frac{1}{2},+}H^{R}_{\frac{1}{2},+}\big)
\Big] \,,\nonumber\\[0.1cm]
\frac{{\rm d}\Gamma^{\lambda_2=-1/2}}{{\rm d}q^2}=&\frac{m_\ell^2}{q^2}\Big[\frac{8}{3}|1+C_{VL}|^2\big(|H^{L}_{-\frac{1}{2},-}|^2\!+\!|H^{L}_{-\frac{1}{2},0}|^2+
3|H^{L}_{-\frac{1}{2},t}|^2\big)\!+\frac{8}{3}|C_{VR}|^2\big(|H^{R}_{-\frac{1}{2},-}|^2\!+\!|H^{R}_{-\frac{1}{2},0}|^2+
3|H^{R}_{-\frac{1}{2},t}|^2\big)\!\nonumber\\[0.1cm]
&\hspace{0.4cm}+\!\frac{64}{3}|C_T|^2\big(|H^{T2}_{-\frac{1}{2},+,-}\!+\!H^{T2}_{-\frac{1}{2},0,t}\!|^{2}
+\!|H^{T2}_{-\frac{1}{2},0,-}\!+\!H^{T2}_{-\frac{1}{2},-,t}|^{2}\big)\!\nonumber\\[0.1cm]
&\hspace{0.4cm}+\!\frac{16}{3}\,{\rm Re}[(1+C_{VL})C_{VR}^{*}]\big(H^{L}_{-\frac{1}{2},-}H^{R}_{-\frac{1}{2},-}+H^{L}_{-\frac{1}{2},0}H^{R}_{-\frac{1}{2},0}
+3H^{L}_{-\frac{1}{2},t}H^{R}_{-\frac{1}{2},t}\big)\Big]\nonumber\\[0.1cm]
+&\frac{16m_\ell}{\sqrt{q^2}}\,{\rm Re}\big[[(1+C_{VL})^{*}H^{L}_{-\frac{1}{2},t}+C_{VR}^{*}H^{R}_{-\frac{1}{2},t}](C_{SL}H^{SPL}_{-\frac{1}{2},0}
+C_{SR}H^{SPR}_{-\frac{1}{2},0})
\nonumber\\[0.1cm]
&\hspace{0.8cm}-2C_T^{*}[(1+C_{VL})H^{L}_{-\frac{1}{2},0}+C_{VR}H^{L}_{-\frac{1}{2},0}](H^T_{-\frac{1}{2},+,-}
+H^T_{-\frac{1}{2},0,t})\nonumber\\[0.1cm]
&\hspace{0.8cm}-2C_T^{*}[(1+C_{VL})H^{L}_{-\frac{1}{2},-}+C_{VR}H^{L}_{-\frac{1}{2},-}](H^T_{-\frac{1}{2},0,-}
+H^T_{-\frac{1}{2},-,t})\big]\nonumber\\[0.2cm]
+&\frac{16}{3}\Big[|1+C_{VL}|^2\big(|H^{L}_{-\frac{1}{2},-}|^2+|H^{L}_{-\frac{1}{2},0}|^2\big)
+|C_{VR}|^2\big(|H^{R}_{-\frac{1}{2},-}|^2+|H^{R}_{-\frac{1}{2},0}|^2\big)\!\nonumber\\[0.1cm]
&\hspace{0.4cm}+\frac{3}{2}\big(|C_{SL}|^2|H^{SPL}_{-\frac{1}{2},0}|^{2}
+|C_{SR}|^2|H^{SPR}_{-\frac{1}{2},0}|^{2}\big)+2|C_T|^2\big(|H^{T}_{-\frac{1}{2},+,-}\!+\!H^{T}_{-\frac{1}{2},0,t}|^{2}\!+\!
|H^{T}_{-\frac{1}{2},0,-}\!+\!H^{T}_{-\frac{1}{2},-,t}|^{2}\big)\!\nonumber\\[0.1cm]
&\hspace{0.4cm}+2{\rm Re}[(1+C_{VL})C_{VR}^{*}]\big(H^{L}_{-\frac{1}{2},-}H^{R}_{-\frac{1}{2},-}+H^{L}_{-\frac{1}{2},0}H^{R}_{-\frac{1}{2},0}\big)
+3{\rm Re}\big(C_{SL}C_{SR}^{*}\big)H^{SPL}_{-\frac{1}{2},0}H^{SPR}_{-\frac{1}{2},0}\Big]\nonumber\\[0.1cm]
\frac{{\rm d}\Gamma^{\lambda_\tau=1/2}}{{\rm d}q^2}=&\frac{m_\ell^2}{q^2}\Big[|1+C_{VL}|^2\big[\frac{8}{3}\big(|H^{VL}_{\frac{1}{2},+}|^2\!+\!|H^{VL}_{\frac{1}{2},0}|^2
\!+\!
|H^{VL}_{-\frac{1}{2},-}|^2\!+\!|H^{VL}_{-\frac{1}{2},0}|^2\big)\!+\!8\big(|H^{VL}_{\frac{1}{2},t}|^2\!
+\!|H^{VL}_{-\frac{1}{2},t}|^2\big)\big]\nonumber\\[0.1cm]
&\hspace{0.9cm}+|C_{VR}|^{2}\big[\frac{8}{3}\big(|H^{R}_{\frac{1}{2},+}|^2\!+\!|H^{R}_{\frac{1}{2},0}|^2\!+\!
|H^{R}_{-\frac{1}{2},-}|^2\!+\!|H^{R}_{-\frac{1}{2},0}|^2\big)\!+\!8\big(|H^{R}_{\frac{1}{2},t}|^2\!
+\!|H^{R}_{-\frac{1}{2},t}|^2\big)\big]\nonumber\\[0.1cm]
&\hspace{0.9cm}+\frac{16}{3}\,{\rm Re}[(1+C_{VL})C_{VR}^{*}]\big(H^{L}_{-\frac{1}{2},-}H^{R}_{-\frac{1}{2},-}+H^{L}_{-\frac{1}{2},0}H^{R}_{-\frac{1}{2},0}
+3H^{L}_{-\frac{1}{2},t}H^{R}_{-\frac{1}{2},t}\nonumber\\[0.1cm]
&\hspace{4.2cm}+H^{L}_{\frac{1}{2},0}H^{R}_{\frac{1}{2},0}+H^{L}_{\frac{1}{2},+}H^{R}_{\frac{1}{2},+}+
3H^{L}_{\frac{1}{2},t}H^{R}_{\frac{1}{2},t}\big)\Big]\nonumber\\[0.1cm]
+&8\Big[|C_{SL}|^2\big(|H^{SPL}_{\frac{1}{2},0}|^{2}+|H^{SPL}_{-\frac{1}{2},0}|^{2}\big)
+|C_{SR}|^{2}\big(|H^{SPR}_{-\frac{1}{2},0}|^{2}+|H^{SPR}_{\frac{1}{2},0}|^{2}\big)\nonumber\\[0.1cm]
&\hspace{0.1cm}+2\,{\rm Re}\big(C_{SL}C_{SR}^{*}\big)\big(H^{SPL}_{-\frac{1}{2},0}H^{SPR}_{-\frac{1}{2},0}+
H^{SPL}_{\frac{1}{2},0}H^{SPR}_{\frac{1}{2},0}\big)
\Big]\nonumber\\[0.1cm]
+&\frac{32|C_T|^2}{3}\Big(|H^{T}_{-\frac{1}{2},-,t}\!+\!H^{T}_{-\frac{1}{2},0,-}|^{2}+|H^{T}_{\frac{1}{2},0,t}+
H^{T}_{\frac{1}{2},+,-}|^{2}+|H^{T}_{\frac{1}{2},+,0}+H^{T}_{\frac{1}{2},+,t}|^{2}+|H^{T}_{-\frac{1}{2},0,t}
+H^{T}_{-\frac{1}{2},+,-}\!|^{2}\Big)
\nonumber\\[0.1cm]
+&\frac{16m_\ell}{\sqrt{q^2}}\,{\rm Re}\Big[[(1+C_{VL})H^{L}_{-\frac{1}{2},t}+C_{VR}H^{R}_{-\frac{1}{2},t}](C_{SL}^{*}H^{SPL}_{-\frac{1}{2},0}
+C_{SR}^{*}H^{SPR}_{-\frac{1}{2},0})
\nonumber\\[0.1cm]&\hspace{1.4cm}
+[(1+C_{VL})H^{L}_{\frac{1}{2},t}+C_{VR}H^{R}_{\frac{1}{2},t}](C_{SL}^{*}H^{SPL}_{\frac{1}{2},0}
+C_{SR}^{*}H^{SPR}_{\frac{1}{2},0})\Big]\nonumber\\[0.1cm]
-&\frac{32m_\ell}{3\sqrt{q^2}}\,\Big[{\rm Re}[(1+C_{VL})C_{T}^{*}]\big(H^{L}_{\frac{1}{2},0}H^T_{\frac{1}{2},+,-}+
H^{L}_{\frac{1}{2},0}H^T_{\frac{1}{2},0,t}+H^{L}_{\frac{1}{2},+}H^T_{\frac{1}{2},+,0}
+H^{L}_{\frac{1}{2},+}H^T_{\frac{1}{2},+,t}\nonumber\\[0.1cm]
&\hspace{2.6cm}+
H^{L}_{-\frac{1}{2},0}H^T_{-\frac{1}{2},+,-}+H^{L}_{-\frac{1}{2},0}H^T_{-\frac{1}{2},0,t}+
H^{L}_{-\frac{1}{2},-}H^T_{-\frac{1}{2},0,-}+H^{L}_{-\frac{1}{2},-}H^T_{-\frac{1}{2},-,t}\big)\nonumber\\[0.1cm]
&\hspace{0.9cm}+{\rm Re}\big(C_{VR}C_{T}^{*}\big)\big(H^{R}_{\frac{1}{2},0}H^T_{\frac{1}{2},+,-}+
H^{R}_{\frac{1}{2},0}H^T_{\frac{1}{2},0,t}+H^{R}_{\frac{1}{2},+}H^T_{\frac{1}{2},+,0}
+H^{R}_{\frac{1}{2},+}H^T_{\frac{1}{2},+,t}\nonumber\\[0.1cm]
&\hspace{2.6cm}+
H^{R}_{-\frac{1}{2},0}H^T_{-\frac{1}{2},+,-}+H^{R}_{-\frac{1}{2},0}H^T_{-\frac{1}{2},0,t}+
H^{R}_{-\frac{1}{2},-}H^T_{-\frac{1}{2},0,-}+H^{R}_{-\frac{1}{2},-}H^T_{-\frac{1}{2},-,t}\big)\Big],\nonumber\\[0.2cm]
\frac{{\rm d}\Gamma^{\lambda_\tau=-1/2}}{{\rm d}q^2}=&\frac{16}{3}\Big[|1+C_{VL}|^{2}\big(|H^{VL}_{\frac{1}{2},+}|^2\!+\!|H^{VL}_{\frac{1}{2},0}|^2\!
+\!|H^{VL}_{-\frac{1}{2},-}|^2\!+
\!|H^{VL}_{-\frac{1}{2},0}|^2\big)\nonumber\\[0.1cm]
&\hspace{0.4cm}+|C_{VR}|^{2}(|H^{R}_{-\frac{1}{2},-}|^{2}+|H^{R}_{-\frac{1}{2},0}|^{2}+|H^{R}_{\frac{1}{2},0}|^{2}
+|H^{R}_{\frac{1}{2},+}|^{2})\nonumber\\[0.1cm]
&\hspace{0.4cm}+2\,{\rm Re}[(1+C_{VL})C_{VR}^{*}]\big(H^{VL}_{-\frac{1}{2},-}H^{R}_{-\frac{1}{2},-}+H^{VL}_{-\frac{1}{2},0}H^{R}_{-\frac{1}{2},0}
+H^{VL}_{\frac{1}{2},0}H^{R}_{\frac{1}{2},0}+H^{VL}_{\frac{1}{2},+}H^{R}_{\frac{1}{2},+}\big)\Big]\nonumber\\[0.1cm]
+&\,\frac{64m_\ell^2}{3q^2}|C_T|^2\big(|H^{T}_{\frac{1}{2},+,-}\!+\!H^{T}_{\frac{1}{2},0,t}|^{2}\!+
\!|H^{T}_{\frac{1}{2},+,0}\!+\!H^{T}_{\frac{1}{2},+,t}|^{2}\!+\!|H^{T}_{-\frac{1}{2},+,-}\!+\!H^{T}_{-\frac{1}{2},0,t}|^{2}
+|H^{T}_{-\frac{1}{2},0,-}\!+\!H^{T}_{-\frac{1}{2},-,t}|^{2}\big)\nonumber\\[0.1cm]
-&\,\frac{64m_\ell}{3\sqrt{q^2}}\,\Big[{\rm Re}[(1+C_{VL})C_{T}^{*}]\big(H^{L}_{\frac{1}{2},0}H^T_{\frac{1}{2},+,-}+
H^{L}_{\frac{1}{2},0}H^T_{\frac{1}{2},0,t}+H^{L}_{\frac{1}{2},+}H^T_{\frac{1}{2},+,0}
+H^{L}_{\frac{1}{2},+}H^T_{\frac{1}{2},+,t}\nonumber\\[0.1cm]
&\hspace{2.6cm}+
H^{L}_{-\frac{1}{2},0}H^T_{-\frac{1}{2},+,-}+H^{L}_{-\frac{1}{2},0}H^T_{-\frac{1}{2},0,t}+
H^{L}_{-\frac{1}{2},-}H^T_{-\frac{1}{2},0,-}+H^{L}_{-\frac{1}{2},-}H^T_{-\frac{1}{2},-,t}\big)\nonumber\\[0.1cm]
&\hspace{0.9cm}+{\rm Re}\big(C_{VR}C_{T}^{*}\big)\big(H^{R}_{\frac{1}{2},0}H^T_{\frac{1}{2},+,-}+
H^{R}_{\frac{1}{2},0}H^T_{\frac{1}{2},0,t}+H^{R}_{\frac{1}{2},+}H^T_{\frac{1}{2},+,0}
+H^{R}_{\frac{1}{2},+}H^T_{\frac{1}{2},+,t}\nonumber\\[0.1cm]
&\hspace{2.6cm}+
H^{R}_{-\frac{1}{2},0}H^T_{-\frac{1}{2},+,-}+H^{R}_{-\frac{1}{2},0}H^T_{-\frac{1}{2},0,t}+
H^{R}_{-\frac{1}{2},-}H^T_{-\frac{1}{2},0,-}+H^{R}_{-\frac{1}{2},-}H^T_{-\frac{1}{2},-,t}\big)\Big].
\end{align}
Based on above helicities, we have two observables, the $q^2$-dependent longitudinal polarizations of $\Lambda_c$ baryon and $\tau$ lepton, the definitions of which are given respectively as
\begin{align}
&P_L^{\Lambda_c}(q^2)=\frac{{\rm d}\Gamma^{\lambda_2=1/2}/{\rm d}q^2-
    {\rm d}\Gamma^{\lambda_2=-1/2}/{\rm d}q^2}{{\rm d}\Gamma^{\lambda_2=1/2}/{\rm d}q^2+
    {\rm d}\Gamma^{\lambda_2=-1/2}/{\rm d}q^2};\\
&P_L^{\tau}(q^2)=\frac{{\rm d}\Gamma^{\lambda_{\tau}=1/2}/{\rm d}q^2-
    {\rm d}\Gamma^{\lambda_{\tau}=-1/2}/{\rm d}q^2}{{\rm d}\Gamma^{\lambda_{\tau}=1/2}/{\rm d}q^2+
    {\rm d}\Gamma^{\lambda_{\tau}=-1/2}/{\rm d}q^2}\,.
\end{align}
In eqs.(\ref{eq:coefficients1}-\ref{eq:coefficients2}), one finds that there are many quadratic terms of $\cos\theta$. In order to show up these terms, one defines a convexity parameter as
\begin{eqnarray}\label{CF1}
C_F^l(q^2) = \frac{1}{ d \Gamma/d q^2} \frac{\,d^2}{d  \cos^2\theta  \,}\Bigg(\frac{d^2 \Gamma}{dq^2 d\cos\theta}\Bigg).
\end{eqnarray}
\section{Results and Discussions} \label{sec:Results}
\subsection{Input Parameters}
In this section, we shall present the numerical results and discussions only for two cases $\ell^-=\mu^-$ and $\tau^-$. Because the results for the electron are almost same as those of muon mode, we will not discuss the electron case any more. In our calculations, the most important non-perturbative parameters are the hadronic transition form factors that have been specified already in previous section. The other parameters we used are list here as
\begin{eqnarray}\label{para}
M_{\Lambda_b}=5.62\,{\rm GeV}, \,\,
M_{\Lambda_c}=2.29\, {\rm GeV}, \,\,
m_{\tau}=1.78\, {\rm GeV}, \,\,
m_{\mu}=0.106\, {\rm GeV},\nonumber\\
G_F=1.66\times 10^{-5}\,{\rm GeV}^{-2}, \,\,
m_b=4.2\,{\rm GeV}, \,\,
m_c=1.28\,{\rm GeV}, \,\,
\tau_{\Lambda_b}=1.466\,{\rm ps}.
\end{eqnarray}
\subsection{SM Predictions}
\begin{figure}[!htb]
\begin{center}
\includegraphics[width=5cm,height=5cm]{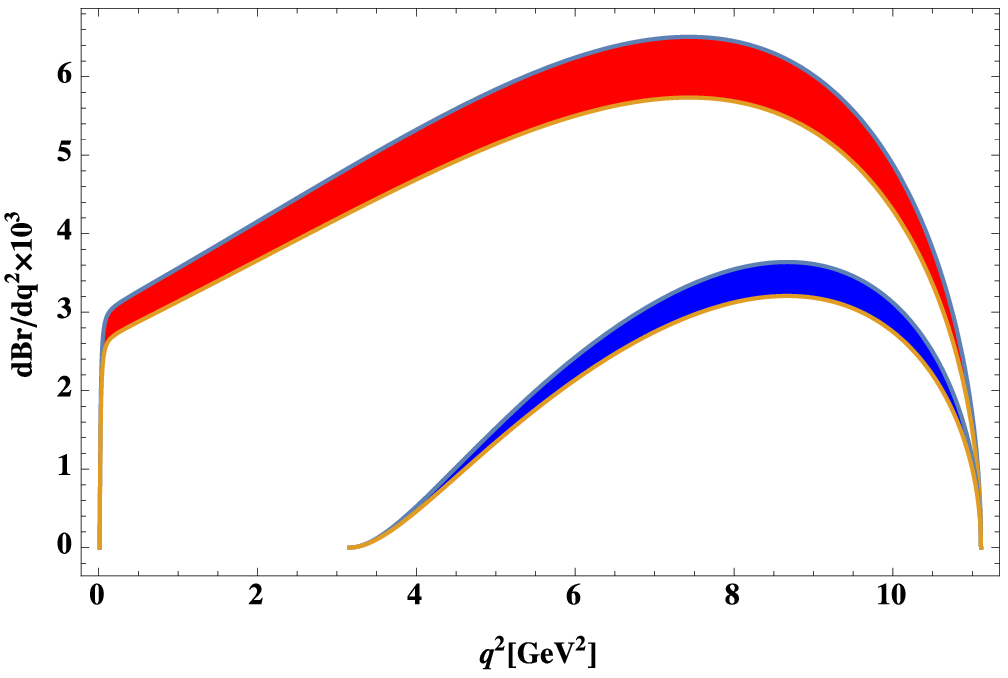}
\includegraphics[width=5cm,height=5cm]{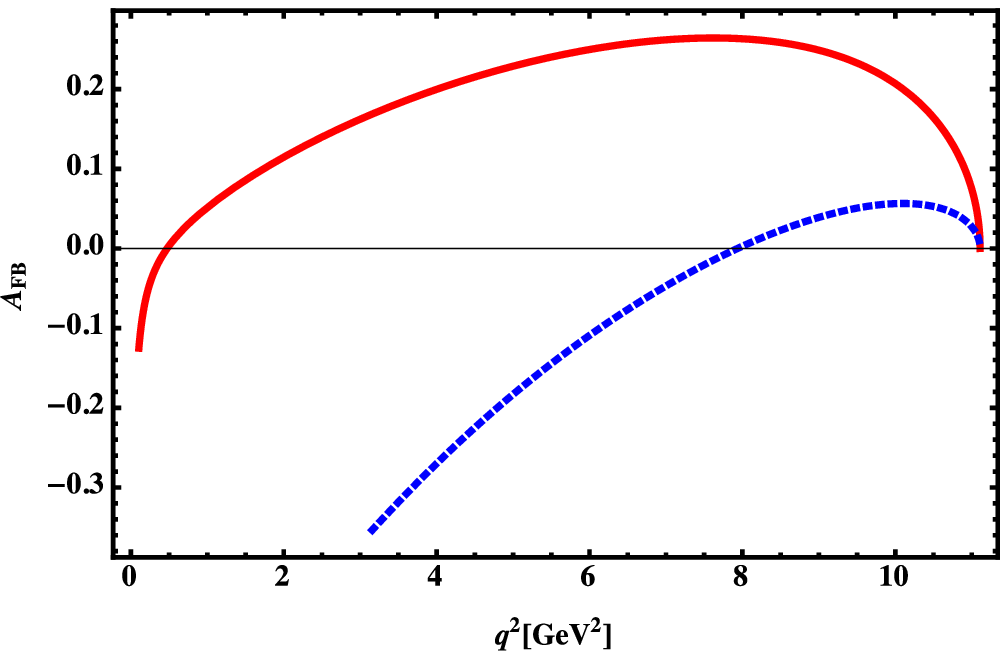}\\
\includegraphics[width=5cm,height=5cm]{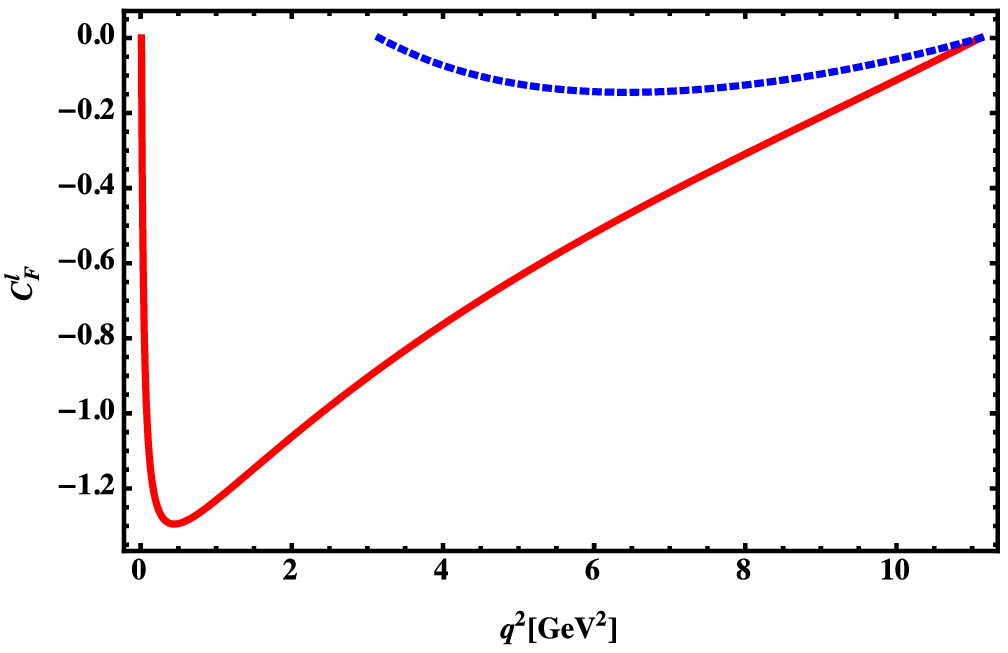}
\includegraphics[width=5cm,height=5cm]{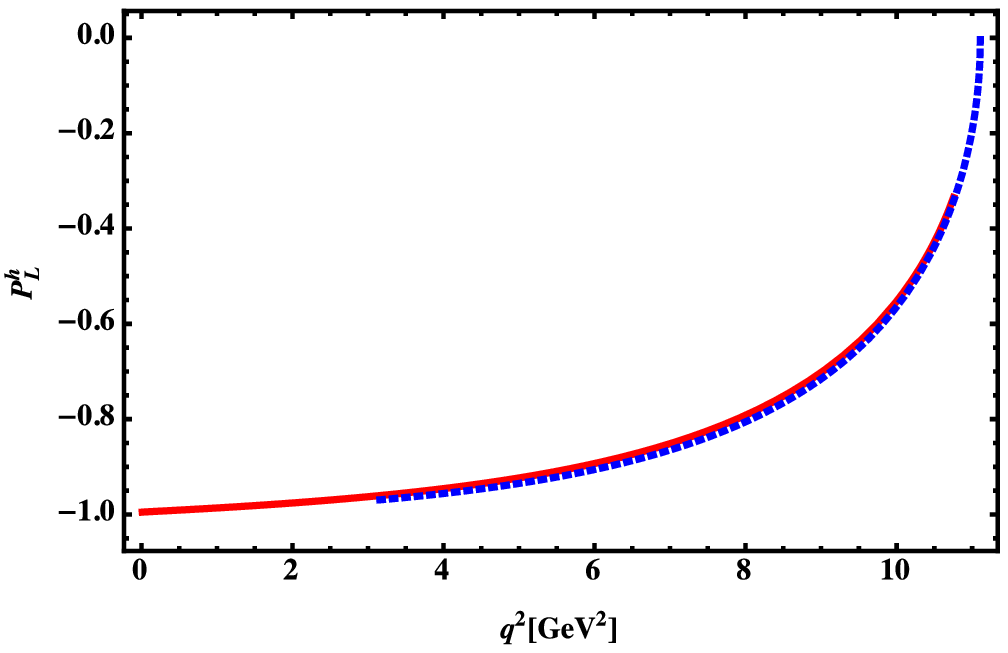}
\includegraphics[width=5cm,height=5cm]{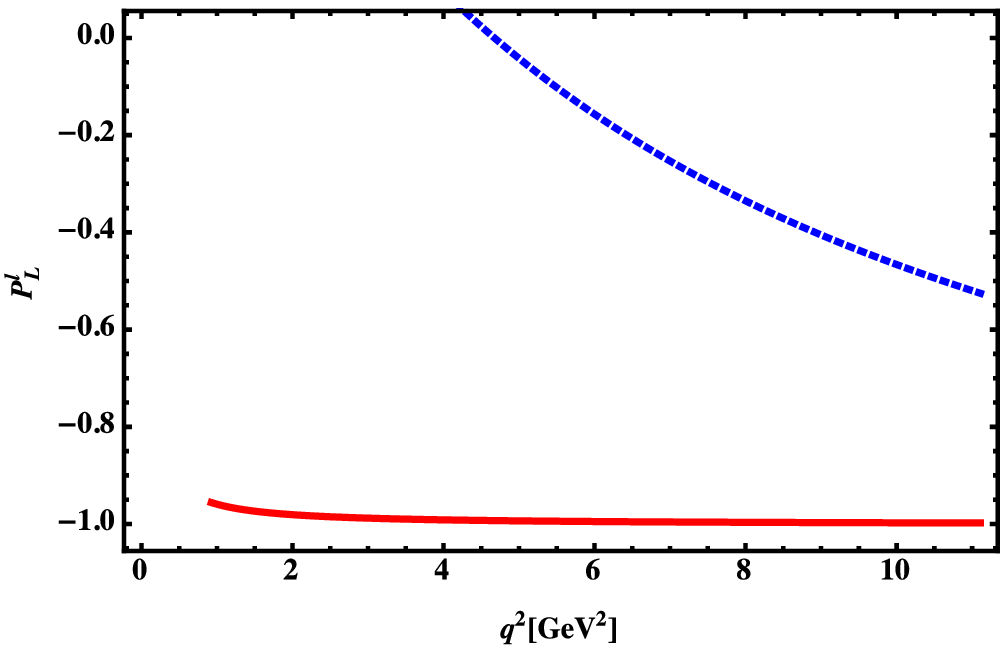}
\caption{The $q^2$-dependence of differential ratios $dBr/dq^2$, the forward-backward asymmetries of the leptonic side $A_{FB}(q^2)$, the convexity parameters $C_F^l(q^2)$, and the longitudinal polarization components of the $\Lambda_c$ and leptons in SM. The red (solid) and blue (dashed) lines indicate muon mode and tau mode, respectively.}\label{Fig:smresults}
\end{center}
\end{figure}

Let us present the numerical results of SM firstly. Using the parameters, we calculate the branching fractions of $\Lambda_b\to\Lambda_c \ell\bar\nu_\ell$ as
\begin{eqnarray}\label{BFSM}
&&{\mathcal B}(\Lambda_b\to\Lambda_c \mu\bar\nu_\mu)|^{\rm SM}=(5.34\pm 0.35)\times 10^{-2};\\
&&{\mathcal B}(\Lambda_b\to\Lambda_c \tau\bar\nu_\tau)|^{\rm SM}=(1.78\pm 0.10)\times 10^{-2};
\end{eqnarray}
and the ratio ${\cal R}(\Lambda_b)$ is given by
\begin{eqnarray}
\mathcal{R}(\Lambda_b)^{\mathrm{SM}}=0.33\pm0.01.
\end{eqnarray}
In Figure.\ref{Fig:smresults}, the $q^2$-dependence of the differential ratios $d\Gamma/dq^2$, the forward-backward asymmetries of the leptonic side $A_{FB}(q^2)$, the longitudinal polarization components of the $\Lambda_c$ and leptons ($P_L^{h}(q^2)$ and $P_L^{l}(q^2)$), and the convexity parameter $C_F^l(q^2)$ are presented, respectively. As seen from Figure.~\ref{Fig:smresults}, for the muon mode, the differential ratio shows a step-like behavior when $q^2=m_{\mu}^2$. At zero recoil $q^2=(M_{\Lambda_b}-M_{\Lambda_c})^2$, $A_{FB}^{\ell}$ approaches to zero for both $\mu$ and $\tau$ modes. However, at large recoil, the $A_{FB}^{\ell}$ are $-0.12$ and $-0.35$, for both $\mu$ and $\tau$ modes, respectively. Furthermore, $A_{FB}^{\mu}$ is positive in most range, and the negative value arises from the small mass of muon. Due to the large mass of $\tau$ lepton, the $A_{FB}^{\tau}$ becomes negative when moving away from zero recoil, and it goes through zero point at $q^2=7.93$ $\rm GeV^{2}$. For the parameter $C_F^{\ell}$, at zero recoil limit, both $C_F^{\mu}$ and $C_F^{\tau}$ are zero. At large recoil range, $C_F^{\mu}=-1.4$ when $q^2=0.4$ $\rm GeV^{2}$, while $C_F^{\mu}$ changes to zero quickly when $q^2=m_{\mu}^2$ due to the effect of mass. This behavior indicates that the $\cos \theta$ distribution in $q^2 \in [0.4,11] \rm GeV^{2}$ is strongly parabolic. On the contrary, the $C_F^{\tau}$ is quite small in the whole range, which implies a straight-line behavior of the $\cos \theta$ distribution. For the longitudinal polarization of the $\Lambda_c$, it is found that at zero recoil the longitudinal polarizations are zero for two modes, while they approach to $-1$ in the large recoil regions. In the last panel, the $q^2$ dependence of the longitudinal polarization of the charged lepton are displayed. For the muon mode, the mass of muon lepton can be neglected in the $\Lambda_b$ decays, thus the muon lepton can be viewed as purely left-handed, and $P_L^{\mu}\simeq -1$ in whole region. While for the tau mode, the behavior is quite different, $P_L^{\tau}= -0.52$ at the zero recoil, and $P_L^{\tau}= 0$ at the maximal recoil point.

\subsection{Model-independence}
\begin{figure}[!htb]
\begin{center}
\includegraphics[scale=0.6]{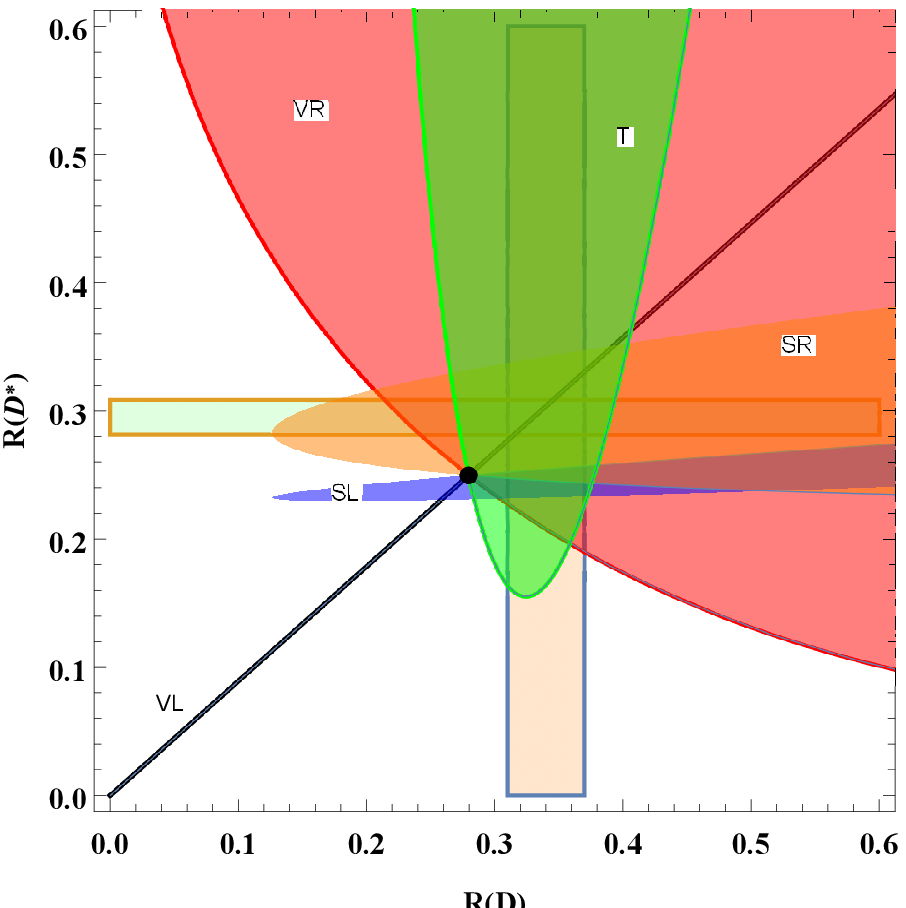} \,\,\,\,
\includegraphics[scale=0.6]{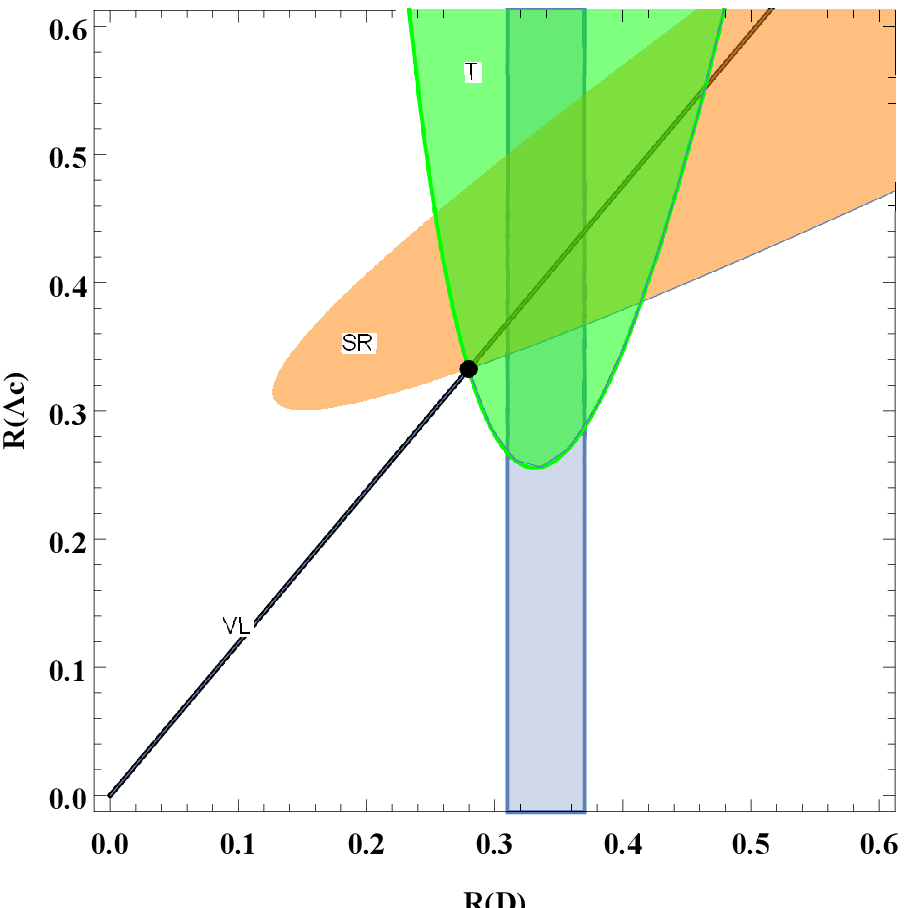}\,\,\,\,
\includegraphics[scale=0.6]{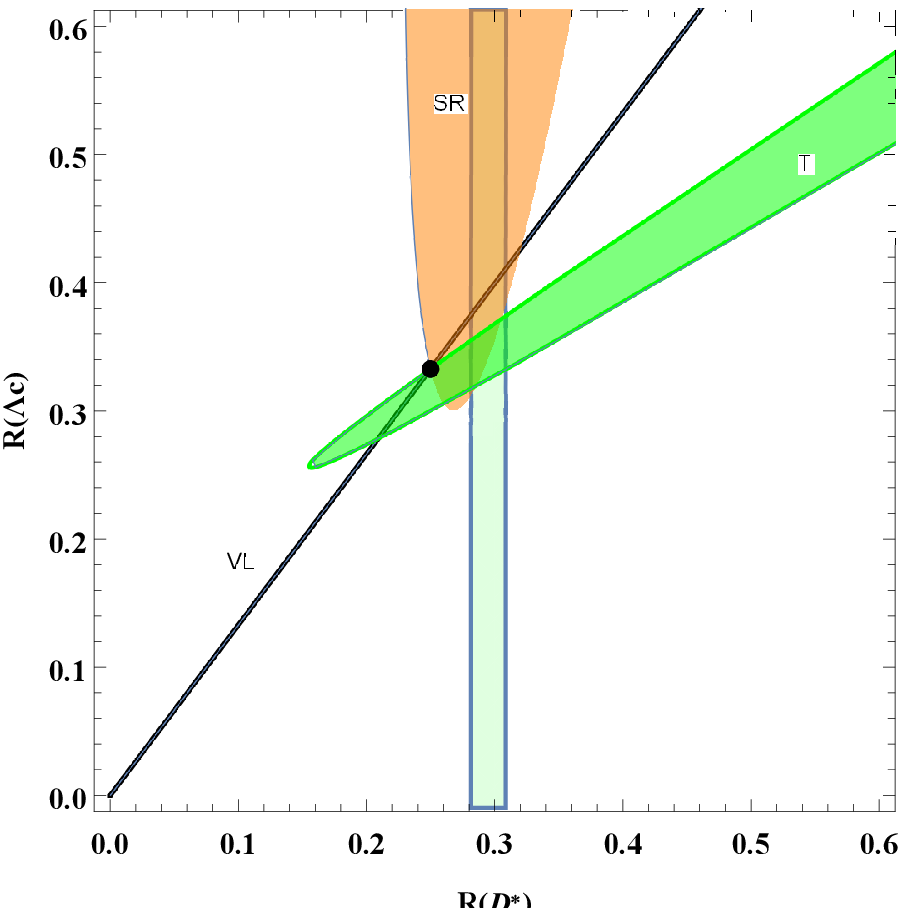}
\caption{Correlations between ${\cal R}(D)$, ${\cal R}(D^{*})$ and ${\cal R}(\Lambda_c)$ in the presence of single NP operators. The vertical and horizontal bands show the experimental constraints, and the black dots denote the SM predictions.}\label{fig:relation}
\end{center}
\end{figure}

We now present our discussions in the model-independence manner. Motivated by tensions stated in previous section, many works have been carried out including the effects of NP, by modifying the wilson coefficients or introducing new operators. As we known, in the quark level, $\Lambda_b \to \Lambda_c \tau^- \bar \nu_{\tau}$, $B\to D^{(*)}\tau^- \bar \nu_{\tau}$, $B_c\to J/\psi (\eta_c)\tau^- \bar \nu_{\tau}$ and $B_c^- \to \tau^- \bar \nu_{\tau}$ are all induced by the $b\to c \tau^- \bar \nu_{\tau}$ transition. So, once we assume that ${\mathcal R}(D^{(*)})$ were induced by NP, it should affect the observables of the $\Lambda_b \to \Lambda_c \tau^- \bar \nu_{\tau}$ decays. In Figure.\ref{fig:relation}, we at first depict the correlations among the ${\mathcal R}(D)$, ${\mathcal R}(D^{*})$ and ${\mathcal R}(\Lambda_c)$ in the presence of each single NP operator given in eq.(\ref{eq:operators}). In the calculations, the form factors of $B\to D^{(*)}$ we employed are from the heavy quark effective theory including higher power and higher order corrections \cite{Huang:2018nnq}. The horizontal and vertical bands represent the experimental constraints at $1\sigma$ confidence level (C.L.)\cite{Amhis:2016xyh}. It is found from the left panel that the ${\cal O}_{VL}$, ${\cal O}_{VR}$, ${\cal O}_{SR}$ and ${\cal O}_{T}$ operators can accommodate the experimental results of ${\mathcal R}(D^{(*)})$ within $1\sigma$, and ${\cal O}_{SL}$ will be ruled out. Furthermore, the ${\cal O}_{VR}$ covers most region, and has less prediction power, so we will not discuss ${\cal O}_{SL}$ and ${\cal O}_{VR}$ temporarily. Using these three figures, we can predict the value of ${\cal R}(\Lambda_c)$ with the existence of one of three operators. For instance, if the new physics only contributes to ${\cal O}_{VL}$ operator, one can read the ${\cal R}(\Lambda_c)\in [0.37, 0.44]$ from the center panel that describes the correlation between ${\cal R}(D)$ and ${\cal R}(\Lambda_c)$. Also, ${\cal R}(\Lambda_c)\in [0.37, 0.41]$ can be arrived based on the relation between ${\cal R}(D^{*})$ and ${\cal R}(\Lambda_c)$. When combining them together, one can predict that ${\cal R}(\Lambda_c)\sim [0.37,0.41]$, which is a bit larger than that of SM. For the operator ${\cal O}_{SR}$, the predicted ${\cal R}(\Lambda_c)$ is in $[0.31,0.45]$, and the main constraint is from ${\cal R}(D)$. On the contrary, for the operator ${\cal O}_{T}$, the constraint arises mainly from the ${\cal R}(D^*)$, and ${\cal R}(\Lambda_c)$ is predicted to be in $[0.28,0.37]$. In this respect, even if we measure ${\cal R}(\Lambda_c)$ is close to the SM prediction, we cannot exclude the contributions from ${\cal O}_{SR}$ and  ${\cal O}_{T}$ operators.

It should be emphasized that in the above discussions, we have not considered the constraints from the recent  measurement of ${\cal R}(J/\psi)$ and the upper limit of branching fraction of pure leptonic decay $B_c^- \to \tau^- \bar \nu_\tau$. In fact, once these constraints were added, all single-operator cannot accommodate the $1\sigma$ experimental constraints any more, which have been concluded in ref.\cite{Huang:2018nnq}. In order to make a general model-independent analysis, the authors in ref.\cite{Huang:2018nnq} had performed a minimum $\chi^2$ fit of the wilson coefficients to the experimental data of all related observables, namely ${\cal R}(D)$, ${\mathcal R}(D^{*})$, ${\cal R}(J/\psi)$, the longitudinal polarizations of $\tau$ lepton and final meson $(D^*)$, with single-operator assumption. As aforementioned, the Belle collaboration released their new results of ${\cal R}(D^{(*)})$ \cite{Abdesselam:2019dgh} at Moriond EW 2019. So, within the latest data, we shall refit the wilson coefficients including $2\sigma$ uncertainties, and the updated results are listed in Table.\ref{fitresults}. The technological details are refereed to ref.\cite{Huang:2018nnq}. It is noted that in contrast to ref.\cite{Huang:2018nnq} where all experimental results from Babar, Belle and LHCb were input, we here only input the five averaged results. With the new obtained wilson coefficients, we recalculate ${\cal R}(D)$, ${\mathcal R}(D^{*})$ and  ${\cal R}(\Lambda_c)$, and they are also presented in Table.\ref{fitresults}, from which one finds that all operators can enhance the ${\cal R}(\Lambda_c)$. Typically, for the operator $\cal{O}_{SL}$, the ${\cal R}(\Lambda_c)$ can be as large as $0.44$.

\begin{table}[!htb]
\begin{center}
\caption{Fitted values of the wilson coefficients with different NP operator.}
\begin{tabular}{ccc c c c}
\hline
\hline
NP scenario  &
value        &
$\chi^2/dof$ &
${\mathcal R}(D)$
&${\mathcal R}(D^*)$
&${\mathcal R}(\Lambda_c)$ \\
\hline
$C_{VL}$ &$(1+Re[C_{V_1}])^2+(Im[C_{V_1}])^2=1.20 $   &$1.51/3$ &0.37&0.30&0.40 \\
$C_{VR}$ &$-0.002 \pm0.45i $   &$1.13/3$ &0.36&0.31&0.40\\
$C_{SL}$ &$-0.92 \pm  0.97i $   &$0.91/3$&0.42&0.24&0.44\\
$C_{SR}$ &$0.25$   &$3.40/3$ &0.43&0.25&0.39\\
$C_{T}$ &$-0.005 \pm  0.09i$   &$2.95/3$&0.31&0.30&0.37\\
\hline
SM&&&0.30&0.25&0.33\\
\hline
\hline
\end{tabular}\label{fitresults}
\end{center}
\end{table}
\begin{figure}[!htb]
\begin{center}
\includegraphics[scale=0.55]{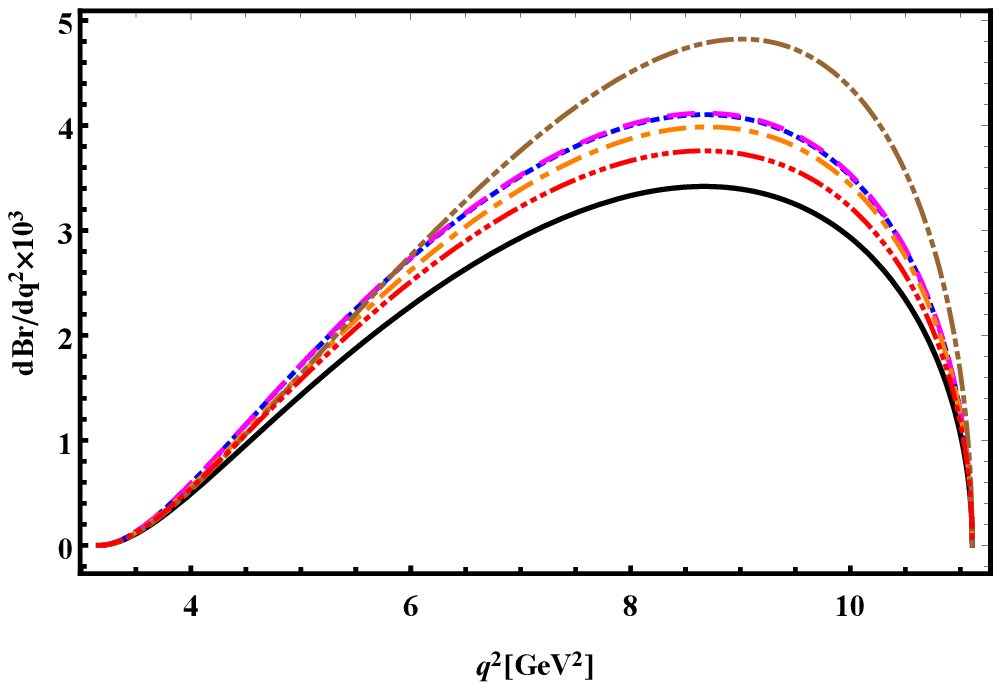} \,\,\,\,
\includegraphics[scale=0.55]{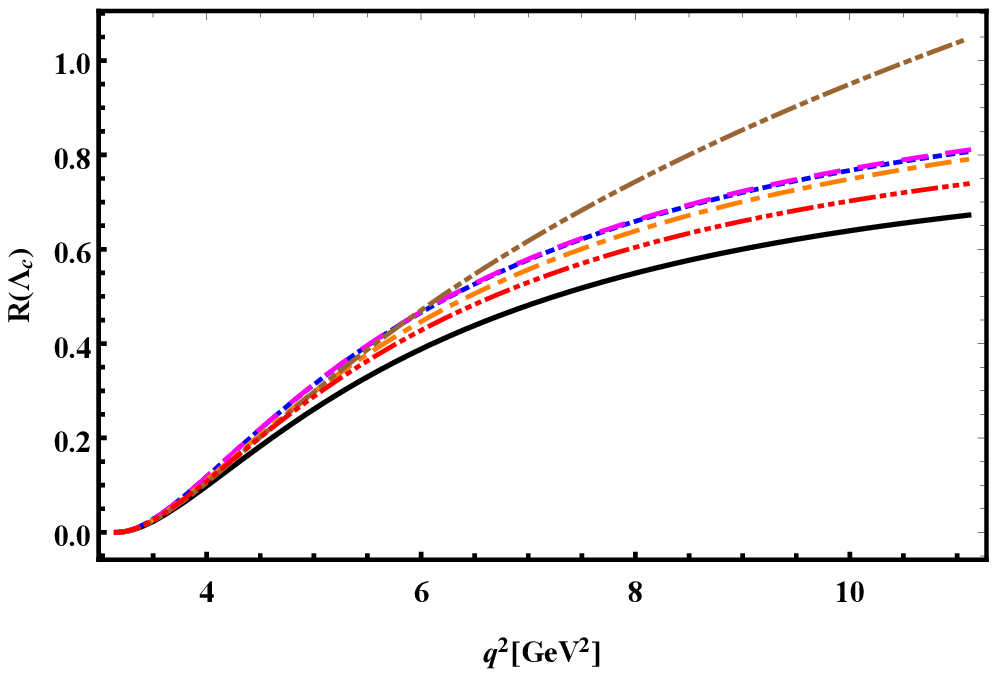} \,\,\,\,
\caption{Predictions for the differential ratio  of $\Lambda_b\to\Lambda_c\tau\bar\nu_\tau$ and ${\cal R}(\Lambda_c)$. The solid (black) lines denotes the SM predictions. The dotted (blue), dashed (magenta), dot-dashed (orange), double-dot dashed (brown) and tridot-dashed (red) lines mean NP predictions corresponding to the best-fit Wilson coefficients of ${\cal O}_{VL}$, ${\cal O}_{VR}$, ${\cal O}_{SL}$, ${\cal O}_{SR}$, and ${\cal O}_{T}$, respectively.}\label{Figure-4}
\end{center}
\end{figure}
\begin{figure}[!htb]
\begin{center}
\includegraphics[scale=0.6]{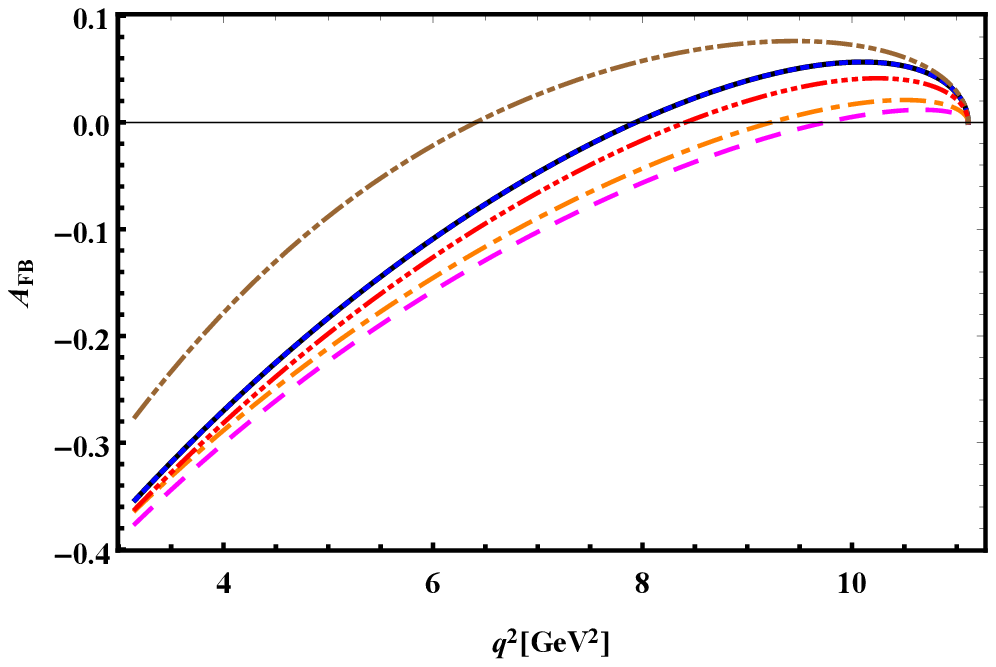} \,\,\,\,
\includegraphics[scale=0.6]{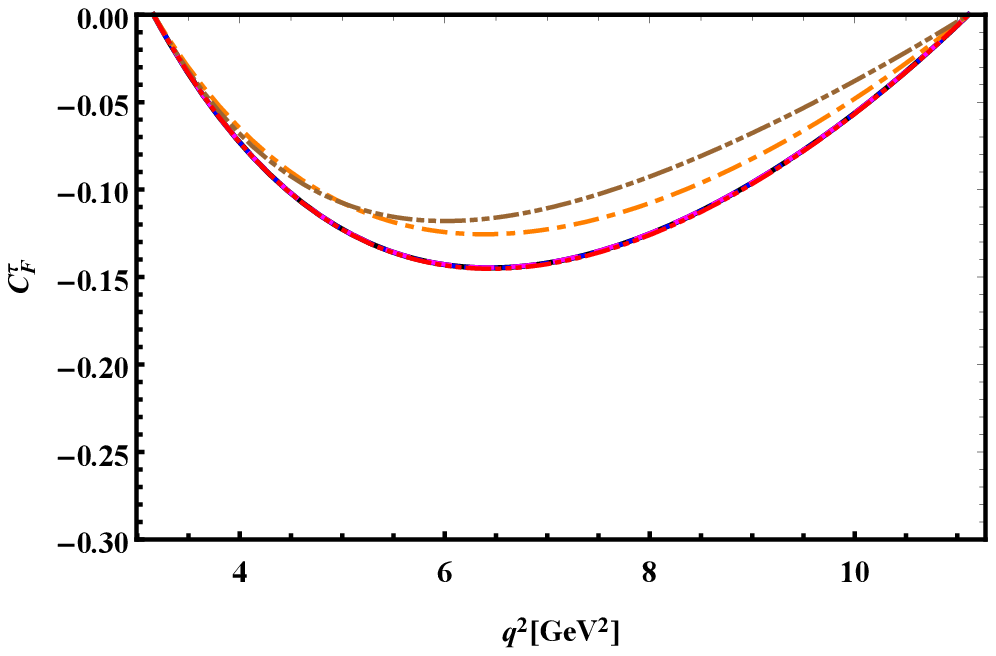} \,\,\,\,\\
\includegraphics[scale=0.6]{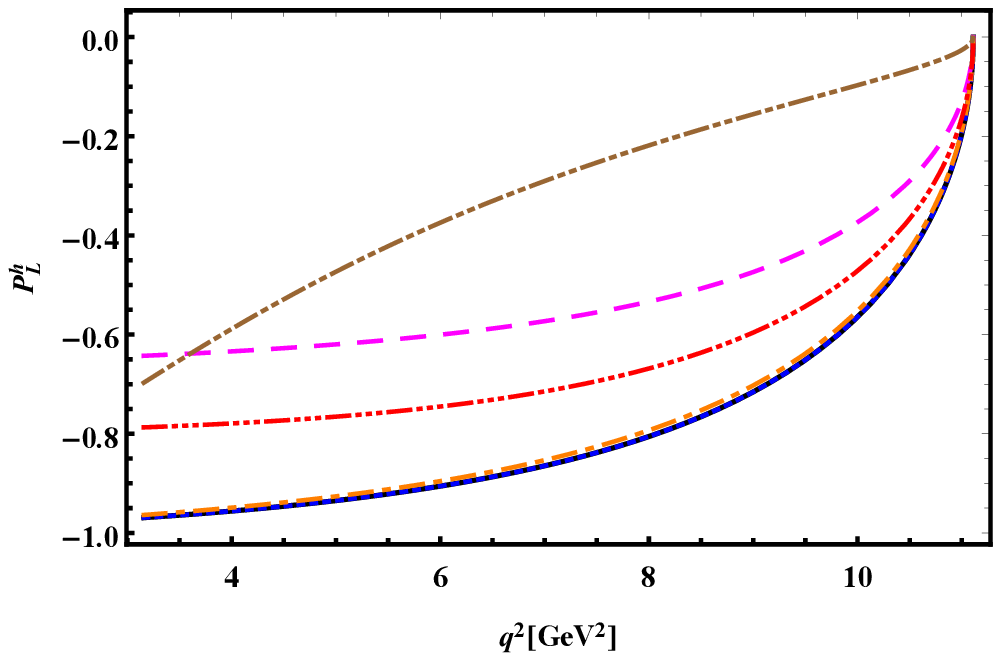} \,\,\,\,
\includegraphics[scale=0.6]{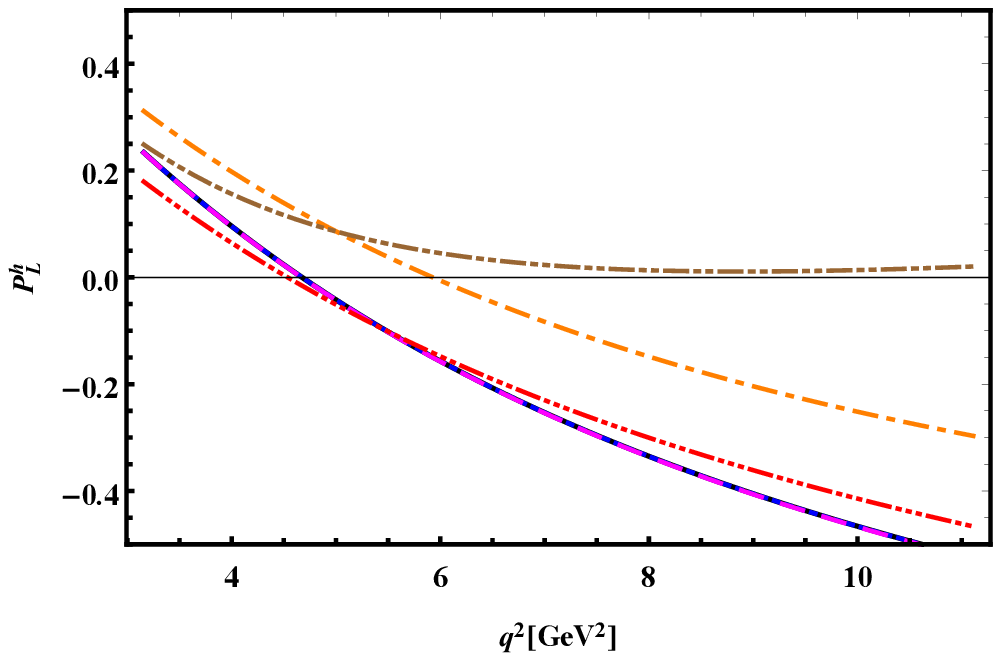} \,\,\,\,
\caption{Predictions for the forward-backward asymmetry, the convexity parameter $C_F^l(q^2)$, the longitudinal polarization components of the $\Lambda_c$ and leptons $P_L^{h}(q^2)$. The solid (black) lines  denote the SM predictions. The dotted (blue), dashed (magenta), dot-dashed (orange), double-dot dashed (brown) and tridot-dashed (red) lines mean NP predictions corresponding to the best-fit Wilson coefficients of ${\cal O}_{VL}$, ${\cal O}_{VR}$, ${\cal O}_{SL}$, ${\cal O}_{SR}$, and ${\cal O}_{T}$,  respectively.}\label{Figure-5}
\end{center}
\end{figure}

With the new wilson coefficients, we also plot the $q^2$-dependence of the differential branching fraction ${\rm d}\mathcal B(\Lambda_b\to\Lambda_c\tau\bar\nu_\tau)/{\rm d}q^2$ and the ratio ${\cal R}(\Lambda_c)$ in Fig.~\ref{Figure-4}. It is can be seen that these two observables are sensitive to the NP operators. Moreover, the effects of the $O_{SL}$ operator are remarkable, and ones of other four operators are very similar, although all of them can enhance the branching fraction. Furthermore, we also show in Fig.~\ref{Figure-5} the $q^2$-dependence of the $\Lambda_c$ and $\tau$ longitudinal polarizations as well as the lepton-side forward-backward asymmetry and the convexity parameter $C_F^l(q^2)$. For the ${\cal O}_{VL}$ operator, since the resulting wilson coefficients $(1+C_{VL})$ appears  in both the numerator and the denominator of $P_L^{\Lambda_c,\tau}$ and $A_{FB}^\tau$, the contributions of NP effects are cancelled out by each other. As for other operators, due to the interference between SM and NP, the effects of NP can show up. The effects of ${\cal O}_{SL}$ are significant. To be honest, there are certain uncertainties in the hadronic form factors. However, because the uncertainties from the form factor can be removed largely, these four diagrams can be used to be the probes of the new physics, if there observables can be measured precisely in future, such as in LHCb or the proposing circular electron-positron collider.

\subsection{Models}
In this section, we shall discuss some typical NP models that could affect $b \to c \tau\bar \nu_\tau$ transition. As we stated previously, the intention of measurements of ${\cal R}(D^{(*)})$ is to search for the charged Higgs-like scalars, such as $H^\pm$ in 2HDMs. As known well, the charged Higgs boson contributes $b \to c \tau\bar \nu_\tau$ at tree level and its effects is enhanced in some 2HDMs by introducing ${\cal O}_{SL}$ and ${\cal O}_{SR}$ operators. Unfortunately, these models are disfavored with the combined experimental results, see ref.\cite{Lees:2012xj}. Besides that, models with charged Higgses lead to unexpected large effects in the total $B_c$ lifetime, and can also be disfavored by the $q^2$ distribution. In these respects, we will not discuss the Higg-like models in this current work. Since the charged Higgs effects on $b \to c \tau\bar \nu_\tau$ in the minimal supersymmetric standard model are the same as those in the 2HDM of type II at the tree level, we will not study it, either. As for the $R$-parity violating minimal supersymmetric standard model, it may affect $b \to c \tau\bar \nu_\tau$ via mediating the slepton and down-squark. However, because its behavior is very similar to the leptoquarks that we will discuss later, we will not discuss this model any more. In the following, we mainly focus on the $W^\prime$ model, the scalar leptoquark (LQ) model and the vector LQ model as well.

\subsubsection{$W^\prime$ Model}
One simple way to obtain a new physics contribution to $b \to c \tau\bar \nu_\tau$  is to use $W^\prime$ gauge boson which couples to the second- and third- generation fermions \cite{Greljo:2015mma}. Without the right-handed neutrinos,  the general Lagrangian that describes the couplings of an extra $W^\prime$ boson to quarks and leptons can be written as \cite{Gomez:2019xfw}
\begin{eqnarray} \label{LagWprime}
\mathcal{L}_{\rm eff}^{W^\prime} = \frac{W^\prime_\mu}{\sqrt{2}}  \Big[\bar{u}_i \gamma^\mu(\epsilon^L_{u_id_j} P_L + \epsilon^R_{u_id_j} P_R) d_j + \bar{\ell}_i  \epsilon^L_{\ell_i\nu_{j}}   \gamma^\mu P_L\nu_{j}\Big] + {\rm h.c.},
\end{eqnarray}
with $u_i \in (u,c,t)$, $d_j \in (d,s,b)$ and $\ell_i, \ell_j \in (e,\mu,\tau)$.  $\epsilon^{L,R}_{u_id_j}$ and $\epsilon^L_{\ell_i\nu_{j}}$ are the dimensionless coupling parameters that can be constrained by the current data of direct or indirect searches. In addition, for simplicity, all the effective couplings are set to be real. Because we hope the new physics only appear in the $\tau$ sector, so only the matrix element $\epsilon^L_{\tau \nu_\tau}$ is assumed to be nonzero. If we transfer the eq.(\ref{LagWprime}) to the effective Lagrangian as eq.(\ref{eq:lag}), the wilson coefficients including the contribution of $W^\prime$ can be expressed as
\begin{eqnarray} \label{CLLW}
C_{VL} \equiv  \dfrac{\sqrt{2}}{4G_F V_{cb}} \dfrac{\epsilon_{cb}^{L} \epsilon_{\tau \nu_\tau}^{L}}{M_{W^\prime}^2},  \,\,\,\,\,\,
C_{VR} \equiv \dfrac{\sqrt{2}}{4G_F V_{cb}}  \dfrac{\epsilon_{cb}^{R} \epsilon_{\tau\nu_\tau}^{L}}{M_{W^\prime}^2},
\end{eqnarray}
where $M_{W^\prime}$ is the mass of the ${W^\prime}$ boson, and $\epsilon_{cb}^{L,R}$ and $\epsilon_{\tau \nu_\tau}^{L}$ are the flavor-dependent couplings of eq.~\eqref{LagWprime}. As aforementioned, the ${\cal O}_{VR}$ cannot contribute to LFU violation in the first order within SMEFT, so we only discuss the effect of the left-handed $W^\prime$ currently. In fact, when all current experimental constraints are taken into account, the permitted region of $|\epsilon_{cb}^{R} \epsilon_{\tau\nu_\tau}^{L}|/{M_{W^\prime}^2}$ is barely reduced, though the authors of ref. \cite{Bhattacharya:2019olg} stated that the right-handed $M_{W^\prime}$ could induce large CP asymmetry in $B \to D^*\mu\nu $ decay mode.

\begin{figure}[!htb]
\begin{center}
\includegraphics[scale=0.7]{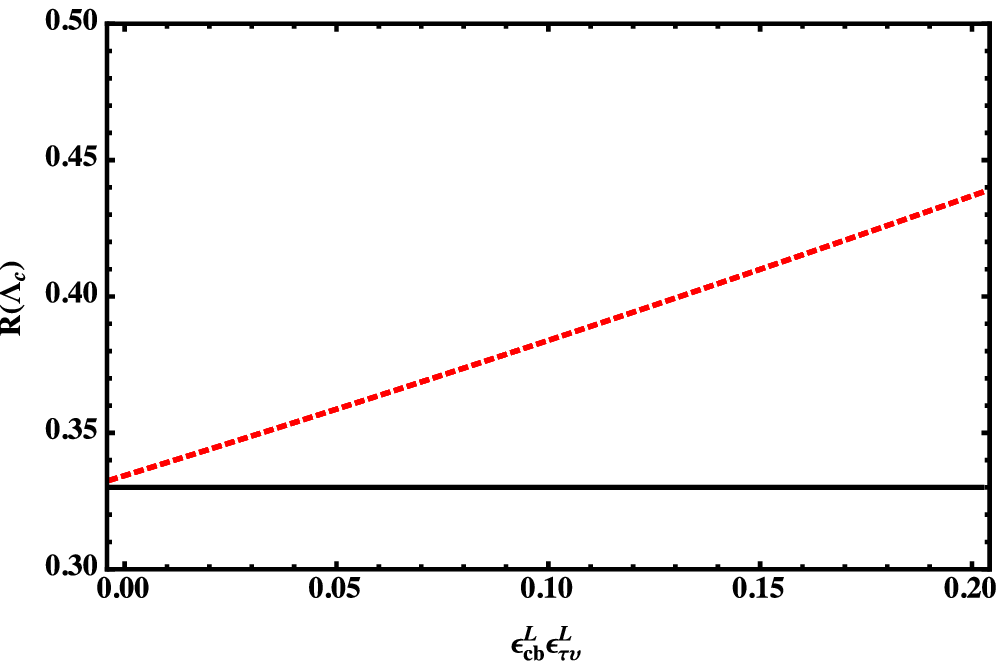}
\includegraphics[scale=0.7]{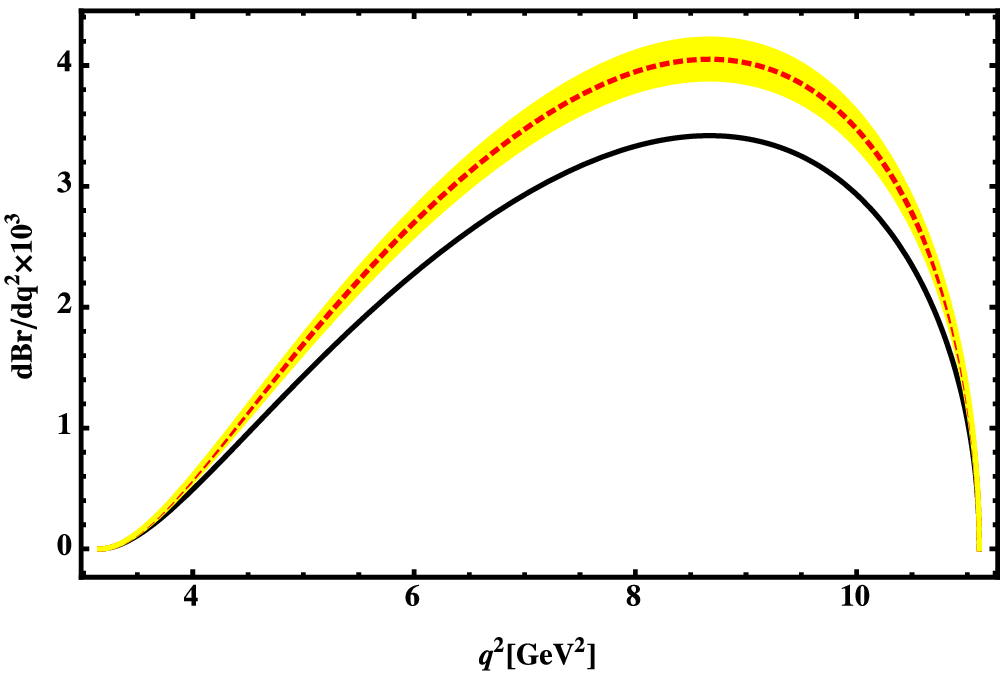} \,\,\,\,\\
\caption{The prediction of ${\cal R}(\Lambda_c)$ as a function of $\epsilon_{cb}^{L} \epsilon_{\tau \nu_\tau}^{L}$ (left panel), and $q^2$ variation of the differential branching fraction (right panel) in the left-handed $W^\prime$ model. The black and red lines indicate the results of SM and the left-handed $W^\prime$ model, respectively. The yellow band is from the uncertainties of $W^\prime$ model.}\label{Figure-6}
\end{center}
\end{figure}

To constrain the parameters in eq.(\ref{CLLW}), many works have been preformed by analysing the ${\cal R}(D^{(*)})$ anomalies, such as in ref.\cite{Greljo:2018tzh, Iguro:2018fni, Abdullah:2018ets,Gomez:2019xfw}. In ref.~\cite{Greljo:2018tzh}, within the current data of ATLAS~\cite{Aaboud:2018vgh} and CMS~\cite{Sirunyan:2018lbg} the authors studied $R(D^{(\ast)})$ anomalies and the mono-tau signature at the LHC in the left-handed $W^\prime$ model and obtained $\epsilon_{cb}^{L} \epsilon_{\tau \nu_\tau}^{L}=(0.14 \pm 0.03)$ with $W^\prime$ mass in the range $[0.5, 3.5]$ TeV, which was consistent with the value $\epsilon_{cb}^{L} \epsilon_{\tau \nu_\tau}^{L} = 0.107$ obtained in \cite{Iguro:2018fni}. Very Recently, in ref.\cite{Gomez:2019xfw}, the authors performed a $\chi^2$ analysis within the latest experimental data and got the best fitted values $\epsilon_{cb}^{L} \epsilon_{\tau \nu_\tau}^{L} = 0.11$ with $ M_{W^\prime}=1 \rm TeV$, which also agrees with previous studies. So, based on the above discussions, in order to maximize the effects of NP, we adopt the range as
\begin{equation}
\epsilon_{cb}^{L} \epsilon_{\tau \nu_\tau}^{L} = (0.12 \pm 0.03) \bigg(\frac{M_{W^\prime}}{{\rm TeV}} \bigg)^{2}.
\end{equation}
With above parameters, ${\cal R}(\Lambda_c)$ is predicted to be $0.38 \pm 0.03$. Furthermore, we show ${\cal R} (\Lambda_c)$ as a function of $\epsilon_{cb}^{L} \epsilon_{\tau \nu_\tau}^{L}$ in Figure.\ref{Figure-6}, where the $q^2$ dependence of the differential branching fraction is also presented. It is found that the curves with an extra $W^\prime$ boson are on the above of ones from SM. Since we here only introduce a left-handed $W^\prime$ boson, it only affects the wilson coefficient $C_{VL}$ and cannot produce other operators, so that other observables, such as $A_{FB}$, $P_L^{h,l}$ and $C_F^l$, are unchanged, compared with results of SM. In addition, because we here suppose $W^\prime$ only couples to the third family fermions, it cannot be produced in the current proton-antiproton collider, so it can escape from the current constrain of direct searches from ATLAS and CMS at LHC.

\subsubsection{Scalar Leptoquark Model}
\begin{figure}[!htb]
\begin{center}
\includegraphics[scale=0.7]{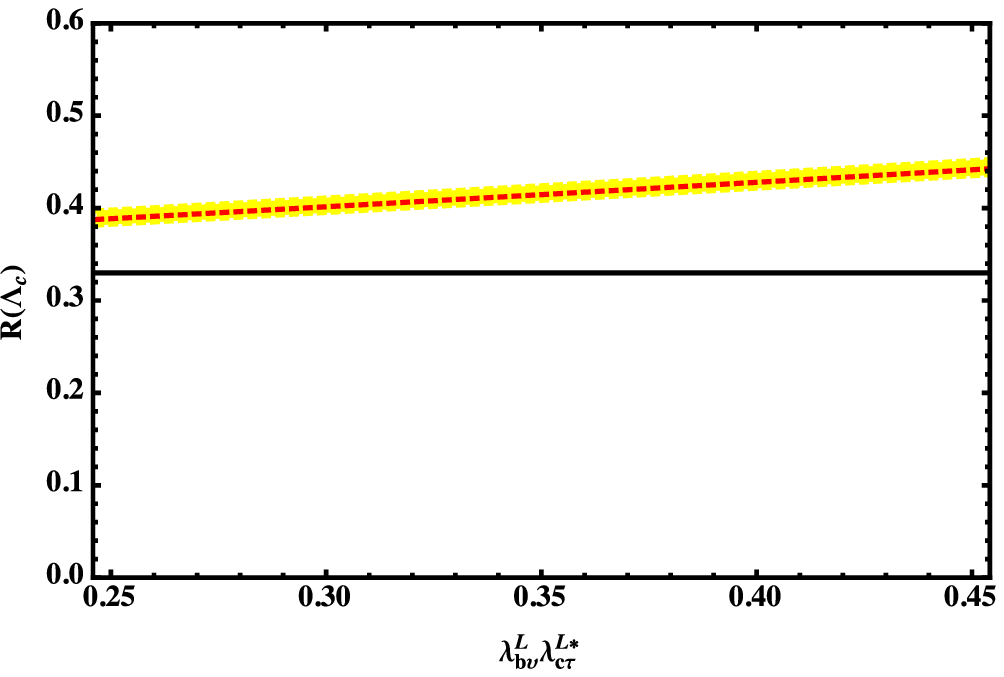}
\includegraphics[scale=0.7]{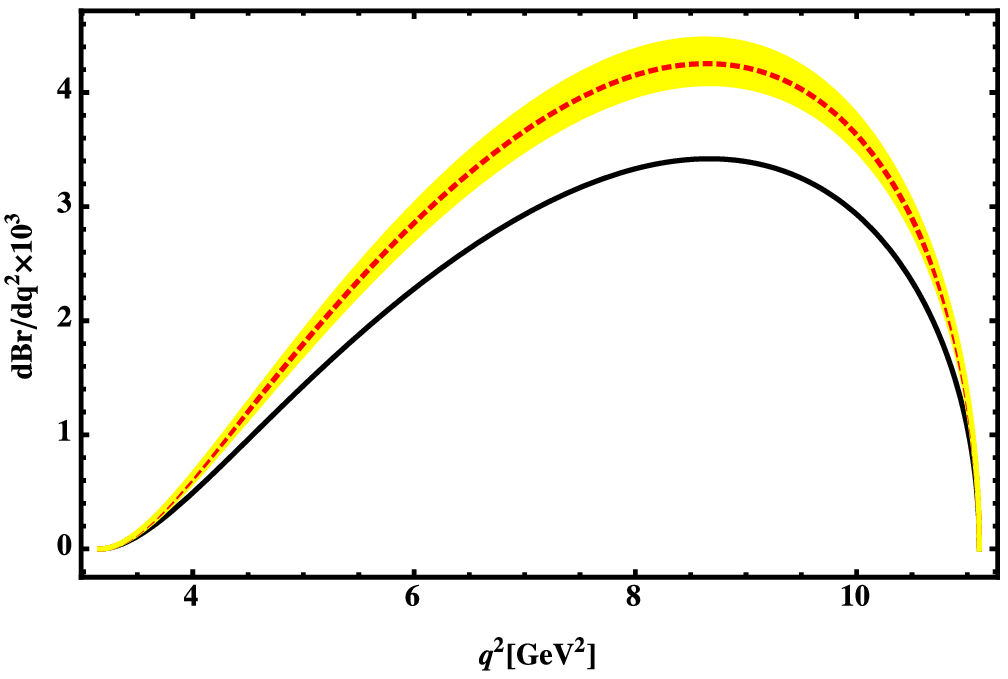} \,\,\,\,\\
\caption{The prediction of ${\cal R}(\Lambda_c)$ as a function of  $\lambda_{b\nu_{\tau}}^L\lambda_{c\tau}^{L\ast}$ (left panel) and the differential branching fraction as a function of $q^2$ (right panel) with a scalar leptoquark. The black and red lines indicate the results of SM and the scalar leptoquark model, respectively. The yellow band is from the uncertainties of NP model. }\label{Figure-7}
\end{center}
\end{figure}

For the $b\to c \ell \bar \nu_\ell$, another potential tree-level mediators are leptoquaks. In the literatures, there are about ten possible leptoquark models that respect the SM symmetry, assuming the leptoquark be a scalar, a vector or a tensor particle. For the scalar particle, we here discuss a model \cite{Bauer:2015knc}, where the LQ $\phi$ transforms under the SM gauge group as $(\mathbf{3}, \mathbf{1}, -\frac{1}{3})$. It is in this model that not only the scalar and vector operators but the tensor operator can be generated. The part of the Lagrangian relevant to the couplings between LQ and SM fermions is given by \cite{Bauer:2015knc}
\begin{equation}\label{eq:scalar coupling}
\mathcal L_{\rm int}^\phi \supset \bar Q_L^c{\boldsymbol\lambda}^Li\tau_2L\phi^{\ast}+\bar u_R^c{\boldsymbol\lambda}^R\ell_R\phi^{\ast}+{\rm h.c.}\,,
\end{equation}
where $Q_L,\,L$ denote the left-handed quark and lepton doublet,  and $u_R,\,\ell_R$ denote the right-handed up-type quark and lepton singlet, respectively. ${\boldsymbol \lambda}^{L,R}$ are the Yukawa coupling matrices in flavour space. More detailed discussions can be found in ref.\cite{Bauer:2015knc}. Such a scalar LQ $\phi$ could induce the $b\to c\tau\bar\nu_\tau$  transition with operators ${\cal O}_{VL}$, ${\cal O}_{SL}$ and ${\cal O}_{T}$, and the corresponding wilson coefficients at scale $\mu=M_\phi$ are given by
\begin{align}
C_{VL}(M_\phi)=\frac{\lambda_{b\nu_{\tau}}^L\lambda_{c\tau}^{L\ast}}
{4\sqrt{2}G_FV_{cb}M_{\phi}^2},\,\,\,\,\,\,
C_{SL}(M_\phi)=-\frac{\lambda_{b\nu_{\tau}}^L\lambda_{c\tau}^{R\ast}}
{4\sqrt{2}G_FV_{cb}M_{\phi}^2},\,\,\,\,\,\,\,
C_T(M_\phi)=-\frac{1}{4}C_{SL}(M_\phi).\label{eq:WCmuphi}
\end{align}
Because the decay we concerned is at scale $\mu=m_b$, all wilson coefficients should be run down to $\mu=m_b$ scale. Due to the conservation of vector current, $C_{VL}$ is not renormalized. In the cases of $C_{SL}$ and $C_T$, the explicit evolution equations and discussions could be found in refs.~\cite{Cai:2017wry,Dorsner:2016wpm}. It should be emphasized that the model with such a scalar LQ can also accommodate the ${\cal R}_K$ and $(g-2)_\mu$ anomalies. Taking $M_\phi = 1 \rm TeV$ as a benchmark, in refs.\cite{Bauer:2015knc, Freytsis:2015qca,Li:2016vvp,Cai:2017wry}, the authors had fitted the parameters within the experimental data, one point they favored is $\lambda_{b\nu_{\tau}}^L \lambda_{c\tau}^{L\ast}=0.35$ and $\lambda_{b\nu_{\tau}}^L \lambda_{c\tau}^{R\ast}=-0.03$. Therefore, to maximize the effect of LQ, we set the parameter space as
\begin{equation}
\lambda_{b\nu_{\tau}}^L\lambda_{c\tau}^{L\ast} = (0.35 \pm 0.05) \bigg(\frac{M_{\phi}}{{\rm TeV}} \bigg)^{2}, \,\,\,\,\,\,
\lambda_{b\nu_{\tau}}^L\lambda_{c\tau}^{R\ast} = (-0.03 \pm 0.03) \bigg(\frac{M_{\phi}}{{\rm TeV}} \bigg)^{2}.
\end{equation}
Using above parameters, we get ${\cal R}(\Lambda_c)=0.41\pm0.02$, which is larger than SM prediction by $25\%$.By setting $M_\phi=1$ TeV, we plot the changes of ${\cal R}(\Lambda_c)$ with $\lambda_{b\nu_{\tau}}^L \lambda_{c\tau}^{L\ast}$ in Figure.\ref{Figure-7}, and  the differential branching fraction as a function of $q^2$ is also displayed.

\subsubsection{Vector Leptoquark Model}
\begin{figure}[!htb]
\begin{center}
\includegraphics[scale=0.7]{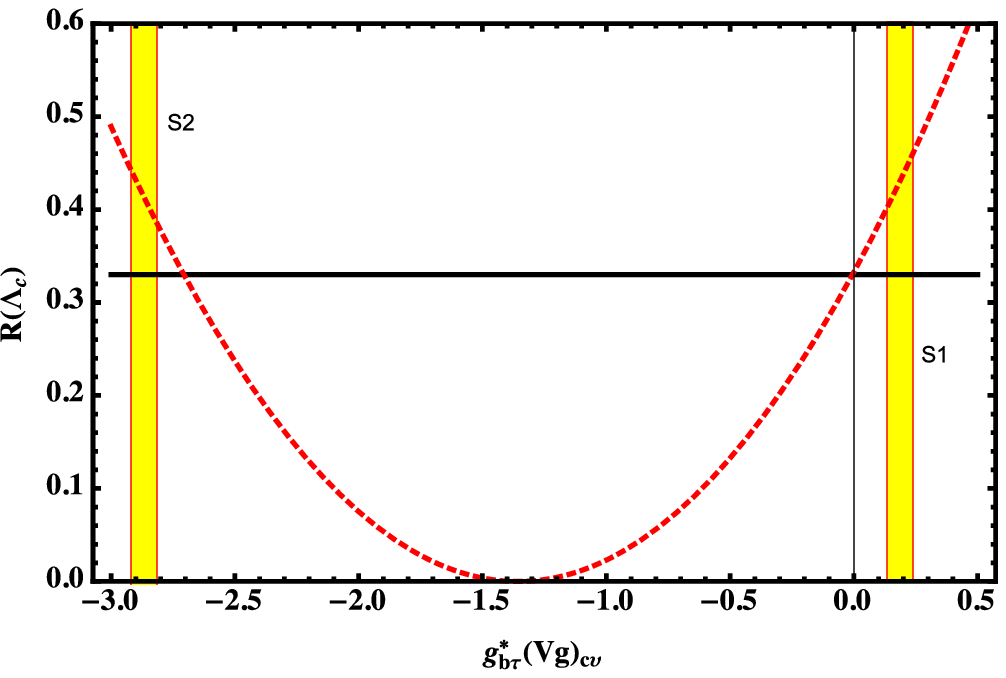} \,\,\,\,
\includegraphics[scale=0.7]{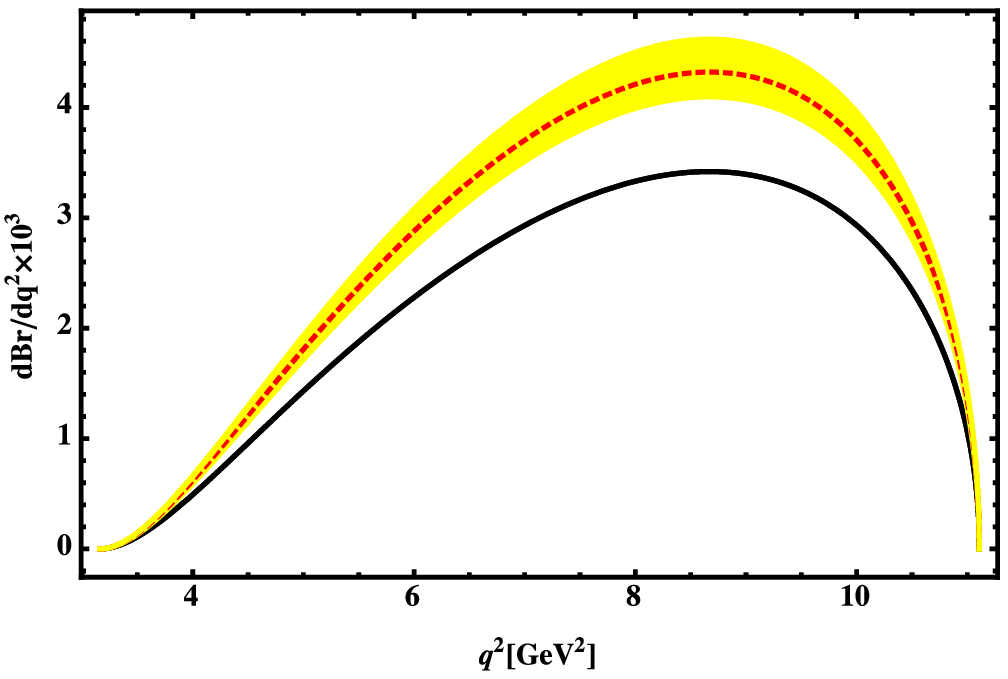} \,\,\,\,\\
\caption{The prediction of ${\cal R}(\Lambda_c)$ as a function of $\lambda_{b\nu_{\tau}}^L\lambda_{c\tau}^{L\ast}$ (left panel) and the differential branching fraction as a function of $q^2$ (right panel) with a vector leptoquark. The black and red lines indicate the results of SM and the vector leptoquark model, respectively. The yellow band is from the uncertainties of NP model. }\label{Figure-8}
\end{center}
\end{figure}

Now, we move to discuss the effect of the vector leptoquark. In this article, we shall explore a model \cite{Fajfer:2015ycq} where the introduced vector $SU(2)_L$ triplet LQ $U_3^\mu$ transforms as $(\mathbf{3}, \mathbf{3},\frac{2}{3})$ under the SM gauge group. In the mass basis, the part of Lagrangian is given by
\begin{align}
 \mathcal L_{U_3} &\supset U_{3\mu}^{(2/3)}\big[(Vg)_{ij}\bar u_i\gamma^\mu P_L\nu_j-g_{ij}\bar d_i\gamma^\mu P_L\ell_j\big]+U_{3\mu}^{(5/3)}(\sqrt2  Vg)_{ij}\bar u_i\gamma^\mu P_L\ell_j \nonumber\\[0.2cm]
 &+U_{3\mu}^{(-1/3)}(\sqrt2g)_{ij}\bar d_i\gamma^\mu P_L\nu_j+{\rm h.c.}\,,
 \end{align}
where $i,j=1,2,3$ denote the $i,j$-th generation fermion. The parameter $g_{ij}$ are the couplings of the $Q=2/3$ component of the triplet, $U_{\mu}^{(2/3)}$, to $\bar d_{Li}$ and $\ell_{Lj}$, and $V$ represents the CKM matrix. From the above Lagrangian, it is found that the $b\to c\tau\bar\nu_\tau$ transition proceeds via exchange of LQ with the operator ${\cal O}_{VL}$. The corresponding wilson coefficient $C_{VL}$ is given correspondingly by
\begin{equation}\label{eq:vector WC}
C_{VL}=\frac{\sqrt2g_{b\tau}^\ast(Vg)_{c\tau}}{4G_FV_{cb}M_U^2}\,.
\end{equation}
It should be noted that in this model, due to the existence of the second term the ${\cal R}(D^{(\ast)})$, ${\cal R}_{K^{(*)}}$ as well as the angular observable ${\cal P}_5^\prime$ in $B\to K^*\mu^+\mu^-$ decay can be explained simultaneously with suitable parameter space, as shown in ref.~\cite{Fajfer:2015ycq}. From the results of $\chi^2$ fits to the measured ratios ${\cal R}(D^{(\ast)})$ and acceptable $q^2$ spectra done in ref. \cite{Freytsis:2015qca}, we learn that at $1\sigma$ we have the following two best-fit solution (S1 and S2)
\begin{align}\label{eq:parameter}
g_{b\tau}^\ast( Vg)_{c\tau} =\bigg(\frac{M_{U}}{{\rm TeV}} \bigg)^{2}
\Bigg\{
      \begin{array}{rl}
        0.18\pm0.04,  & S_1;\\
        -2.88\pm0.04,  & S_2.
      \end{array}
    \Bigg.\,
\end{align}
It can be seen that the vector LQ only affects the functions $A_1$ and $A_2^{VL}$ in eq.(\ref{eq:differential angular}), and both two terms are proportional to $|1+C_{VL}|^2$. Therefore, although the fit results $S_1$ and $S_2$ are quite different, the coefficients $(1+C_{VL})$ they induced have nearly same absolute values with different signs. Currently, it is very hard for us to differentiate these two solutions. With above results, we obtain ${\cal R} (\Lambda_c)=0.43 \pm 0.03$. Also, we plot changes of the ${\cal R}(\Lambda_c)$ with $\lambda_{b \nu_{\tau}}^L \lambda_{c\tau}^{L\ast}$, as well as the differential branching fraction as a function of $q^2$ in Figure.\ref{Figure-8}. From two figures, we find that both the ${\cal R}(\Lambda_c)$ and the branching fractions are enhanced by about $30\%$. From above analyses, one finds that although the left-handed $W^\prime$ and the vector leptoquark particle origin form different models, however for the concerned decay mode, their behaviors are very similar, since both of them only contribute to the operator ${\cal O}_{VL}$.

\section{Summary} \label{sec:Conclusions}
Recent measurements of ratios ${\cal R}(D^{(*)})$ imply that a new physics may appear in the $b\to c \tau \bar\nu_\tau$ transition. So, in this work we explored the effect of NP in $\Lambda_b\to\Lambda_c \tau\bar\nu_\tau$ decay mode that is also induced by $b\to c \tau \bar\nu_\tau$. To show up the NP effect, we firstly calculated the branching fraction  of $\Lambda_b\to\Lambda_c \tau\bar\nu_\tau$ and its ratio to $\Lambda_b\to\Lambda_c \mu\bar\nu_\mu$ in SM, and ${\cal R}(\Lambda_c)=0.33\pm0.01$, within the latest form factors from lattice QCD calculations. When studying the effect of NP, we explore $\Lambda_b\to\Lambda_c \tau\bar\nu_\tau$ in a model-independent manner. The differential and total branching fraction, the longitudinal polarizations of final lepton, the forward-backward asymmetries in the lepton-side, the convexity parameters, and the ratio ${\cal R}(\Lambda_c)$ have been calculated. Supposing that NP only affects the third generation fermion, we also presented the correlations among ${\cal R}(D)$,  ${\cal R}(D^*)$ and ${\cal R}(\Lambda_c)$. Considering the latest experimental data, we perform a minimum $\chi^2$ fit of the wilson coefficient of each operator, and found that the left-handed scalar operator ${\cal O}_{SL}$ could enhance the branching fraction by $30\%$. Using the fitted values of the wilson coefficients, we also give a prognosis for the physical observables of  $\Lambda_b\to\Lambda_c \tau\bar\nu_\tau$, such as the ratio ${\cal R}(\Lambda_c)$, forward-backward asymmetry and other polarized observables as well as the differential branching fraction. At last, we also study the contributions of three typical NP models, the $W^\prime$ model, the scalar leptoquark (LQ) model and the vector LQ model, to the ratio and the differential branching fraction of $\Lambda_b\to\Lambda_c \tau\bar\nu_\tau$. These results can be tested in the current LHCb experiment and the future high energy experiments.

\section*{Acknowledgment}
We thank Prof. X.-Q. Li, Dr. Z.-R. Hang and Dr. C. Wang for valuable discussions. This work was supported in part by the National Natural Science Foundation of China under the Grants No. 11575151,11975195 and 11705159; and the Natural Science Foundation of Shandong province under the Grant No. ZR2018JL001 and No. ZR2016JL001.
\bibliographystyle{bibstyle}
\bibliography{mybibfile}
\end{document}